\documentclass[journal]{IEEEtran}
\usepackage{cite}
\usepackage{amsmath,amssymb,amsfonts}
\usepackage{graphicx}
\usepackage{textcomp}
\usepackage{xcolor}
\usepackage{authblk}

\usepackage{booktabs}
\usepackage{epsfig}
\usepackage{latexsym}
\usepackage{multirow}
\usepackage{stfloats}
\usepackage{epstopdf}
\usepackage{color}  
\usepackage{tabularx} 
\usepackage{enumerate}
\usepackage{array}
\graphicspath{{./Figures/}}
\usepackage{color}
\usepackage{bbm}
\usepackage{caption}
\usepackage{bm}
\usepackage[tight,footnotesize]{subfigure}
\usepackage{balance}
\usepackage{mathrsfs}
\usepackage{verbatim}
\allowdisplaybreaks[4]
\usepackage{dsfont}
\usepackage{verbatim}
\usepackage{tikz}
\usepackage{setspace}
\usepackage{diagbox}
\usepackage[framemethod=tikz]{mdframed}
\usepackage{multicol}
\usepackage{environ}
\usepackage{tikz}
\usepackage{stfloats}
\usepackage{algpseudocode}
\usepackage{graphics}
\usepackage{epsfig}
\usepackage{amsthm}
\usepackage{authblk}
\newcommand{\rrev}[1]{\textcolor{black}{{#1}}}
\newcommand{\rev}[1]{\textcolor{black}{{#1}}}

\ifCLASSOPTIONcompsoc
\usepackage[caption=false, font=normalsize, labelfont=sf, textfont=sf]{subfig}
\else
\usepackage[caption=false, font=footnotesize]{subfig}

\usepackage{bm}
\def\BibTeX{{\rm B\kern-.05em{\sc i\kern-.025em b}\kern-.08em
    T\kern-.1667em\lower.7ex\hbox{E}\kern-.125emX}}
\begin{document}

\title{300~GHz Dual-Band Channel Measurement, Analysis and Modeling in an L-shaped Hallway}

\author{Yiqin Wang, Yuanbo~Li, Chong~Han,~\IEEEmembership{Member,~IEEE}, Yi Chen, and Ziming Yu
\thanks{
This paper was presented in part at the IEEE ICC, 2022~\cite{wang2022thz}.

Yiqin Wang, Yuanbo Li, and Chong Han are with the Terahertz Wireless Communications (TWC) Laboratory, Shanghai Jiao Tong University, Shanghai, China (e-mail: \{wangyiqin, yuanbo.li, chong.han\}@sjtu.edu.cn).

Yi Chen, and Ziming Yu are with Huawei Technologies Co., Ltd, China (e-mail: \{chenyi171, yuziming\}@huawei.com).
}
}



\markboth{\rev{IEEE Transactions on Antennas and Propagation, Submitted in May 2023}}{}
\maketitle
\thispagestyle{empty}
\begin{abstract}
The Terahertz (THz) band (0.1-10~THz) has been envisioned as one of the promising spectrum bands for sixth-generation (6G) and beyond communications. In this paper, a dual-band angular-resolvable wideband channel measurement in an indoor L-shaped hallway is presented and THz channel characteristics at 306-321~GHz and 356-371~GHz are analyzed.
It is found that conventional close-in and alpha-beta path loss models cannot take good care of large-scale fading in the non-line-of-sight (NLoS) case, for which a modified alpha-beta path loss model for the NLoS case is proposed and verified in the NLoS case for both indoor and outdoor L-shaped scenarios.
To describe both large-scale and small-scale fading, a ray-tracing (RT)-statistical hybrid channel model is proposed in the THz hallway scenario.
Specifically in the hybrid model, the deterministic part in hybrid channel modeling uses RT modeling of dominant multi-path components (MPCs), i.e., LoS and multi-bounce reflected paths in the near-NLoS region, while dominant MPCs at far-NLoS positions can be deduced based on the developed statistical evolving model. The evolving model describes the continuous change of arrival angle, power and delay of dominant MPCs in the NLoS region. On the other hand, non-dominant MPCs are generated statistically. The proposed hybrid approach reduces the computational cost and solves the inaccuracy or even missing of dominant MPCs through RT at far-NLoS positions.
\end{abstract}

\section{Introduction}

Rapid growth of the wireless data traffic in the past decades has stimulated the demand for 100 Gigabit-per-second and even Terabit-per-second (Tbps) communications in future wireless applications.
Compared with the millimeter-wave (mmWave) band (30-300~GHz) that has a bandwidth of several gigahertz (GHz), the Terahertz (THz) band (0.1-10~THz), which can support Tbps data rates owing to multi-tens-of-GHz bandwidth, has been envisioned as one of the promising spectrum bands for sixth-generation (6G) and beyond communications~\cite{akyildiz2022terahertz,rappaport2019wireless}. In 2017, IEEE 802.15.3d became the first wireless standard operating at the frequency between 252~GHz and 321~GHz~\cite{802.15.3d}. Furthermore, the frequency bands from 275~GHz to 450~GHz have been identified for the implementation of fixed and land mobile service applications in the World Radiocommunication Conference 2019 (WRC-19)~\cite{WRC-19-FINAL-ACTS}.

When moving up to new spectrum, the foundation of wireless system design is the full knowledge of wireless channels, including understanding of radio propagation, channel characteristics analysis, and channel model development.
On one hand, physical channel measurement supported by a wideband channel sounder is one of the major approaches to investigate THz wireless channels~\cite{han2022terahertz}.
Due to the high path loss and frequency-dependent molecular absorption above 300~GHz, to-date THz channel measurement campaigns, typically, either focused on the ``sub-THz'' band (100-300~GHz)~\cite{chen2021channel,he2021channel,nguyen2021large, abbasi2023thz, ju2022sub, dupleich2020characterization}, or line-of-sight (LoS)~\cite{zantah2019channel, khalid2019statistical, serghiou2020ultra} and short-range (desktop, motherboard, data center)~\cite{kim2016characterization, eckhardt2019measurements} scenarios above 300~GHz.
Though, there are THz channel measurements at and above 300~GHz in a larger-scale environment implemented in a small (about 10~m$^2$) indoor office room~\cite{priebe2011channel}, an urban microcell scenario~\cite{undi2021angle}, railway or vehicular communication scenarios~\cite{guan2021channel,eckhardt2021channel}, indoor hallway scenarios~\cite{wang2022thz,li2022channel}, \rev{a hall scenario~\cite{lyu2023measurement}}, and an aircraft cabin~\cite{doeker2022channel}.
On the other hand, deterministic simulation is another way to study THz wireless channels, which is typically implemented for channel characteristic analysis in the environment where measurement is unavailable~\cite{yang2021analysis}, or for channel data reproduction~\cite{han2022terahertz}.
Among them, deterministic simulators can provide the same channel parameters as the measurement based on geometric theories, and can further provide information on the trajectory of wave propagation.

\rev{In light of the two approaches to study the channel, the ray-tracing (RT)-statistical hybrid channel modeling is envisioned as one promising  methodology towards 6G, in which the deterministic part can be generated by RT simulators, while the statistical part is extracted from measurement data~\cite{chen2021channel}. The combination of deterministic and statistical methods reaches a good balance between the model accuracy and the implementation efficiency.}
However, the accuracy of simulation is vulnerable to the discrepancy between the simulation model and the real world. The situation becomes worse in the THz band, where the material property is inadequate, and especially in the non-line-of-sight (NLoS) case, on account of the long propagation path and the inaccuracy of the scattering model.
Till date, the complete analysis, including the measurement, simulation and hybrid modeling, of the NLoS-region site-scale indoor wireless channel above 0.3~THz is still missing.

In this paper, we first conduct an dual-band angular-resolvable wideband channel measurement in a typical L-shaped hallway at 306-321~GHz and 356-371~GHz. In each frequency band, four points in the LoS case, and fourteen points in the NLoS case are measured. In light of the measurement result, the multi-path components (MPCs) propagation in the L-shaped hallway is traced, and THz channel characteristics in dual bands are analyzed and compared.
Specifically, \rrev{a modified $\alpha-\beta$ path loss model is proposed and verified for the NLoS case in the L-shaped scenario, both indoor and outdoor, since conventional path loss models cannot take care of large-scale fading in the NLoS case.}
Second, we carry out an RT-based simulation and discuss the matching degree between simulation and measurement results, which motivates the modeling of evolution of dominant MPCs in the L-shaped NLoS region.
\rrev{Finally, RT-statistical hybrid channel modeling is tailored to the NLoS hallway scenario by using deterministic RT to generate dominant MPCs at only near-NLoS positions, and then applying the statistical evolving model to deduce the dominant MPCs at far-NLoS positions. The derivation of non-dominant MPCs remains statistical.
}

Compared to our preliminary and shorter version~\cite{wang2022thz}, this work includes not only more measurement efforts by including additional measuring points and one extra frequency band, but also the development and validation of a modified $\alpha-\beta$ path loss model, the analysis and modeling of the evolution of dominant MPCs in the NLoS hallway scenario and the development of a hybrid channel model. The distinctive contributions of this work are summarized as follows.
\begin{itemize}
    \item Extensive measurement in dual bands above 300~GHz is carried out in L-shaped scenarios. Specifically, the angular-resolvable wideband channel measurement investigates eighteen receiver (Rx) positions including LoS and NLoS cases in the L-shaped hallway scenario at 306-321~GHz and 356-321~GHz, and in total 6480 channel impulse responses are obtained. Remarkably, 50~GHz-spaced dual-band measurement is useful to characterize the channel frequency selectivity and guide future communication design.
    \item A modified $\alpha-\beta$ path loss model is proposed and verified for the NLoS case in both indoor and outdoor L-shaped scenarios. Besides, a complete analysis of multi-path propagation and THz channel characteristics in dual bands is provided, and comparison between LoS and NLoS cases and different frequency bands is elaborated. The large-scale fading, delay spread, angular spreads are found statistical-wise same and parameter-wise different between dual bands. The higher frequency band has larger path loss and smaller delay and angular spreads.
    \item An RT-statistical hybrid model is proposed in the L-shaped scenario. 
    The deterministic RT result of dominant MPCs is verified accurate in LoS and near-NLoS positions and yet inaccurate or even missing in far-NLoS positions. In light of this, a statistical evolving model of dominant MPCs is developed, revealing the continuous change of dominant MPCs in the NLoS hallway. The hybrid channel modeling framework is then tailored to the L-shaped NLoS scenario by generating dominant MPCs at near-NLoS positions through deterministic simulation and deducing dominant MPCs at far-NLoS positions with the developed statistical evolving model. On the other hand, non-dominant MPCs are generated statistically. By reusing the RT result at near-NLoS positions, the approach saves the computational cost of channel modeling in the NLoS region.
\end{itemize}

The remainder of this paper is organized as follows.
In Sec.~\ref{section: campaign}, the channel measurement campaign and the data process procedure are introduced.
In Sec.~\ref{section: results}, \rrev{the L-shaped scenario is standardized and} the modified $\alpha-\beta$ path loss model for the NLoS hallway is proposed. Besides, a complete analysis and comparison of multi-path propagation and THz channel characteristics in dual bands is provided.
In Sec.~\ref{section: hybrid_modeling}, \rrev{the RT-based simulation is conducted and the evolution of dominant MPCs is modeled. The RT-statistical hybrid model in the L-shaped scenario is then proposed. }
Finally, the paper is concluded in Sec.~\ref{section: conclusion}.
\section{Channel Measurement Campaign} \label{section: campaign}

In this section, we describe the measurement campaign as well as the data post-processing procedures in the L-shaped hallway. The measurement is carried out on the second floor of the Longbin building, at the University of Michigan-Shanghai Jiao Tong University Joint Institute (UM-SJTU JI), Shanghai Jiao Tong University (SJTU).

\begin{figure}
    \centering
    \includegraphics[width=0.9\linewidth]{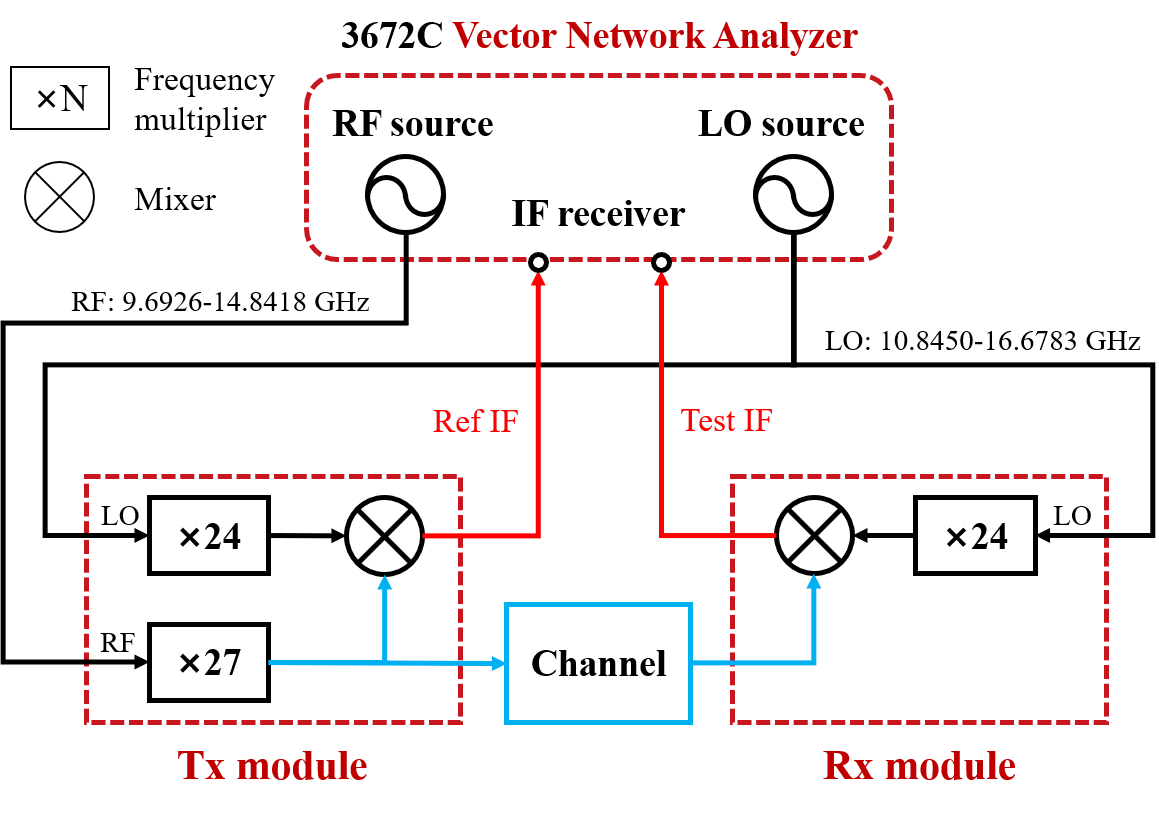}
    \caption{The RF front of the channel measurement system.}
    \label{fig:platform}
\end{figure}

\begin{table}
  \centering
  \caption{Parameters of the measurement system.}
    \begin{tabular}{ll}
    \toprule
    Parameter & Value \\
    \midrule
    Frequency band              & 306-321, 356-371~GHz \\
    Bandwidth                   & 15~GHz \\
    IF frequency                & 7.6~MHz \\
    IF bandwidth                & 1~kHz \\
    Sweeping interval           & 2.5~MHz \\
    Sweeping points             & 6001 \\
    Tx antenna gain / HPBW      & 7~dBi / $60^\circ$ \\
    Rx antenna gain / HPBW      & 25~dBi / $8^\circ$ \\
    Time resolution             & 66.7~ps \\
    Space resolution            & 2~cm \\
    Maximum excess delay        & 400~ns \\
    Maximum path length         & 120~m \\
    Rx azimuth rotation range   & $[0^\circ:10^\circ:360^\circ]$ \\
    Rx elevation rotation range & $[-20^\circ:10^\circ:20^\circ]$ \\
    Average noise floor         & -165~dBm / -180~dBm \\
    Dynamic range               & 65~dB \\
    \bottomrule
    \end{tabular}
  \label{tab:system_parameter}
\end{table}
\begin{figure}
    \centering
    \includegraphics[width=0.9\linewidth]{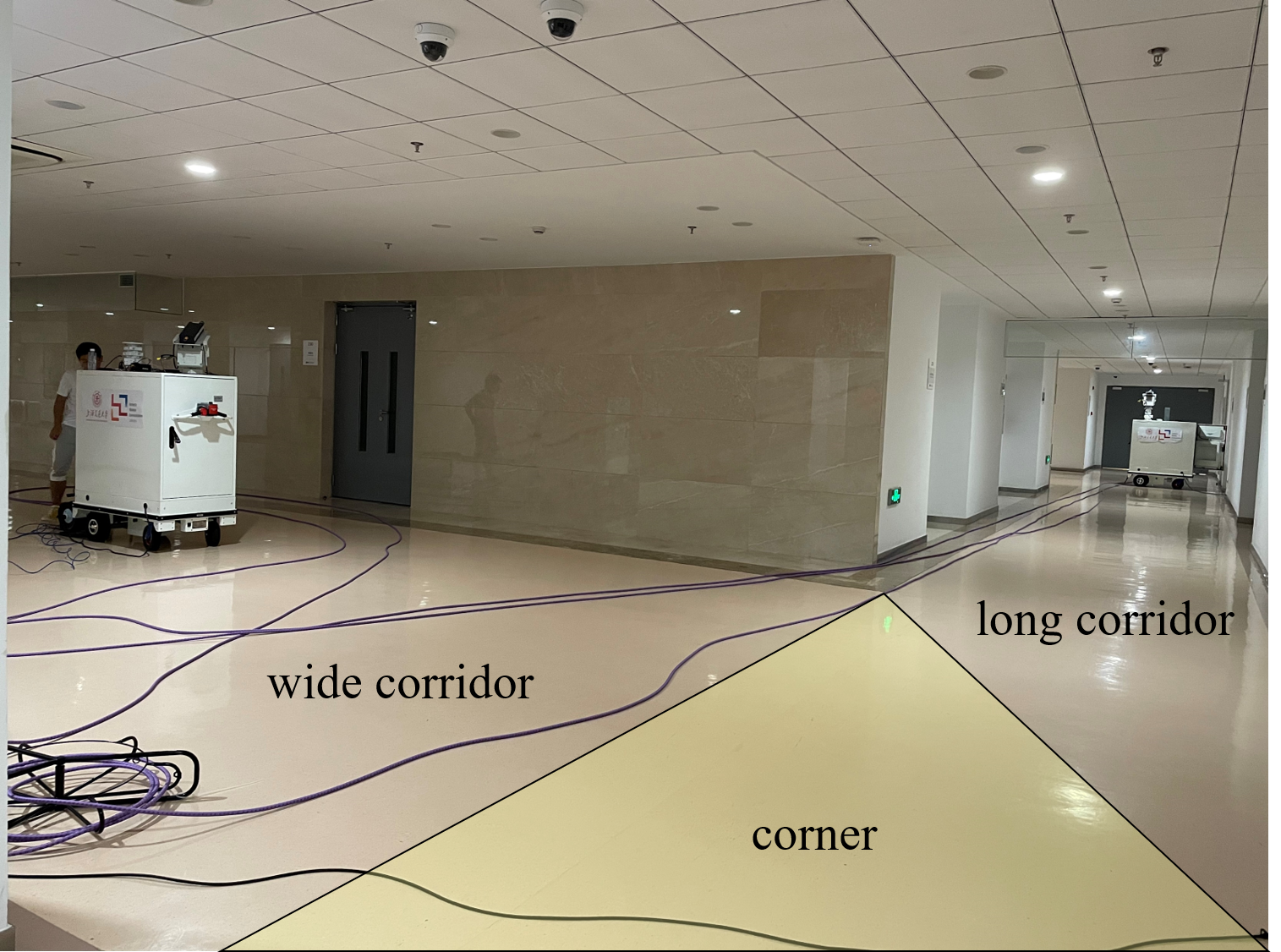}
\caption{The photo of the measurement campaign.}
\label{fig:photo}
\end{figure}

\subsection{Channel Measurement System}
\rev{We adopt the channel measurement platform that consists of radio frequency front ends and a Ceyear 3672C VNA, as illustrated in Fig.~\ref{fig:platform}.}
\rev{The VNA contains the radio frequency (RF) source and the local oscillator (LO) source. The RF signal is multiplied by 27 to reach the target carrier frequency range and directly sent to the wireless channel. The LO signal is multiplied by 24, whose frequency is designed to ensure that the intermediate frequency (IF) signal has the center frequency of 7.6~MHz.}

In this measurement campaign, the dual frequency bands of 306-321~GHz and 356-371~GHz are investigated. \rev{As two sub-bands are separated by 50~GHz, spanning more than 10\% of 300~GHz, they are typically supported by different antenna systems. Therefore, we examine whether the channel characteristics are different in these two sub-bands. For each sub-band,} the measured bandwidth is 15~GHz, which denotes that the time and space resolutions are 66.7~ps and 2~cm, respectively. The frequency sweeping interval is 2.5~MHz, resulting in 6001 sweeping points and a maximum excess delay of 400~ns, which is equivalent to a maximum detectable path length of 120~m. 
The waveguide at the transmitter (Tx) has a gain of around 7~dBi with the half-power beamwidth (HPBW) around $60^\circ$, while the Rx antenna has the gain around 25~dBi and the HPBW of $8^\circ$.

The transceiver module is mounted on a rotator. The Tx is placed on an electric lifter, which is integrated with the electric cart, to reach the height of 2~m above the ground. By contrast, the Rx is put on the top of the cart without a lifter, which results in the height of 1.75~m above the ground. Account for the large HPBW of the Tx antenna, the Tx is static at zero azimuth and elevation angles. By contrast, motored by the rotator, the Rx scans from $0^\circ$ to $360^\circ$ in the azimuth domain and from $-20^\circ$ to $20^\circ$ in the elevation domain to receive multi-path propagation, with the angle step of $10^\circ$, similar to the HPBW of the Rx antenna. The elevation space is not fully investigated since the measuring process is time-consuming. Key parameters of the measurement system are summarized in Table~\ref{tab:system_parameter}.

\begin{figure}
    \centering
    \subfigure[The definition of an L-shaped scenario.]{
    \includegraphics[width=\linewidth]{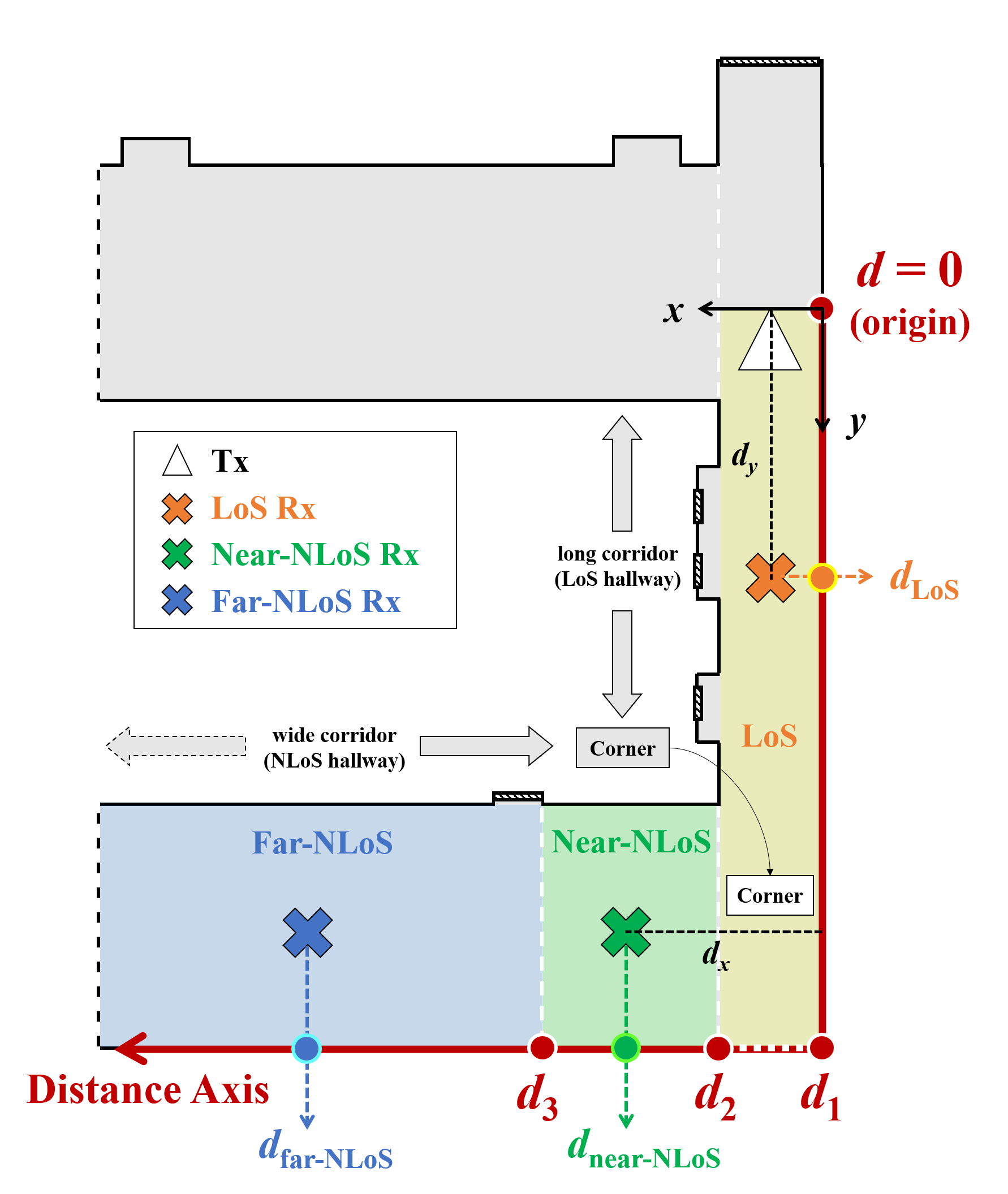}
    }
    \\
    \subfigure[Measurement deployment (unit: mm).]{
    \includegraphics[width=\linewidth]{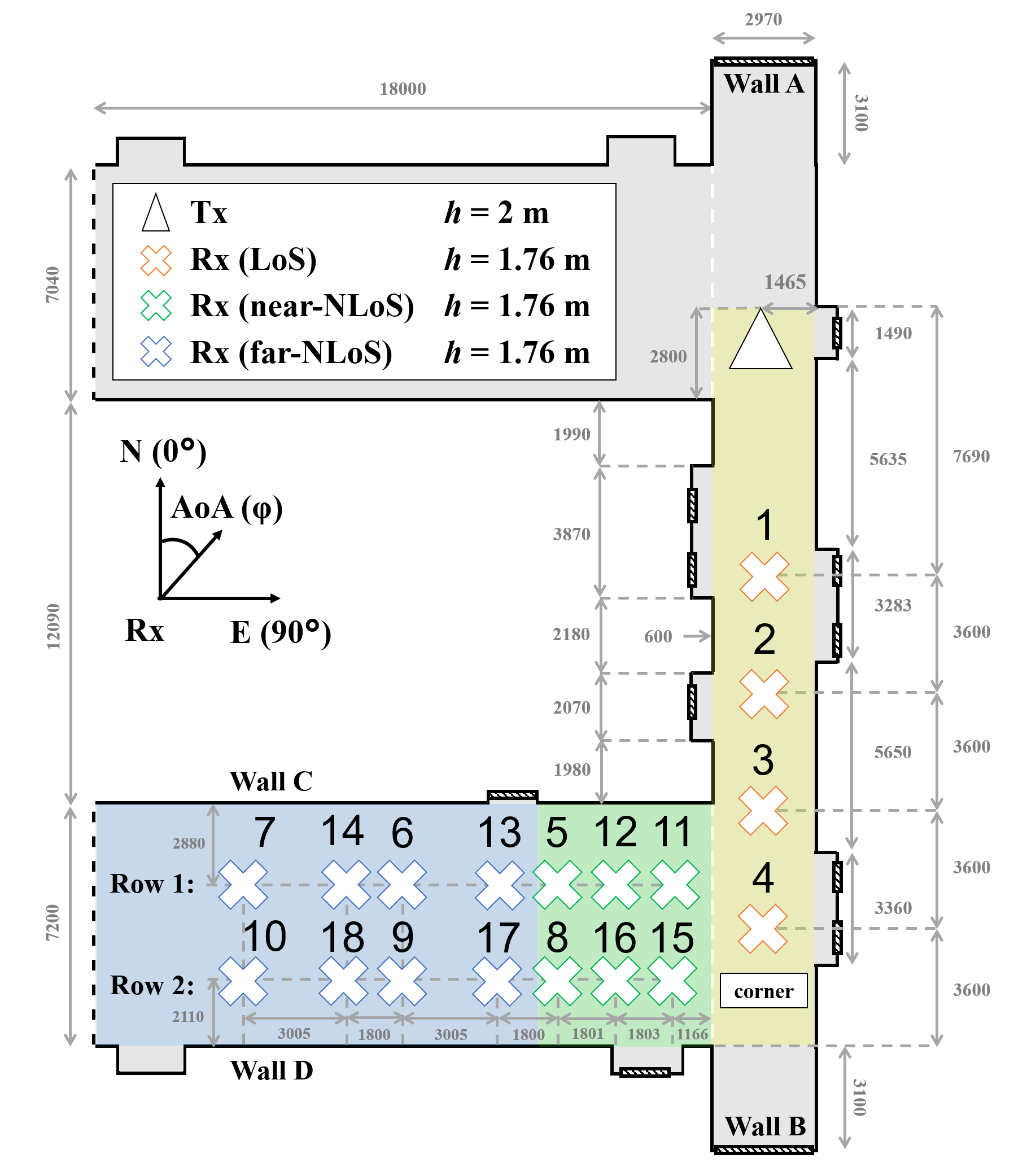}
    }
    \caption{Measurement deployment in an L-shaped hallway, 2$^{\rm nd}$ floor of UM-SJTU JI building, SJTU.}
    \label{fig:deployment}
\end{figure}

\subsection{Measurement Deployment}
The measurement campaign is conducted in the L-shaped hallway on the second floor of the building of UM-SJTU JI, as shown in Fig.~\ref{fig:photo}. \rrev{In Fig.~\ref{fig:deployment}(a), an indoor L-shaped scenario is defined as two perpendicular corridors connected by a corner. Depending on the position of Tx, Rx in two corridors are in the LoS and NLoS case, respectively. Therefore, we abbreviate the two cases to the LoS hallway and the NLoS hallway for simplicity. Furthermore, we use a special distance axis along the outer edge of the L-shape. Instead of the Tx-Rx straight-line distance, we characterize distance as the path length along the bent axis from the origin to the projection of the Rx onto the axis. The projection of the Tx onto the axis is set as the origin. Moreover, three particular values are defined. $d_{1}$ denotes the end of the LoS case, $d_{2}$ is the start of the NLoS case, and $d_{3}$ divides near-NLoS and far-NLoS regions. The region between $d_{1}$ and $d_{2}$ is not defined in this model and is determined by the width of the LoS hallway. If ($d_{x}$, $d_{y}$) represents the traditional Cartesian coordinate of each Rx position, then $d=d_y$ for LoS positions, and $d=d_x+d_1$ for NLoS positions.}

The measurement deployment is illustrated in Fig.~\ref{fig:deployment}(b). The campaign mainly consists of two perpendicular corridors, connected by a corner. The long corridor is 2.97~m wide and 32.53~m long (between Wall~A and Wall~B), including two 3~m long extensions at both ends. Indented offices are distributed along the corridor, whose depth of indent is about 0.6~m with doors closed. The other corridor is 7~m wide (between Wall~C and Wall~D) which extends all the way to the left end.
Tx is placed in the long corridor (the LoS hallway), together with four LoS Rx positions, while fourteen Rx positions are located in the wide corridor (the NLoS hallway). In particular, NLoS Rx are divided into six near-NLoS and eight far-NLoS Rx.
In Section~\ref{section: results} and Section~\ref{section: hybrid_modeling}, different results on the power distribution in the measurement and the existence of the multi-bounce reflected paths in the simulation are discovered between these two kinds of NLoS Rx.
Besides, signals are too weak to be detected in the NLoS hallway far from the corner, so the rest of the corridor to the left end is omitted in Fig.~\ref{fig:deployment}(b).

\subsection{System Calibration and MPC Extraction} \label{sec:clustering_original} 
Before data processing, system calibration is carried out to eliminate the effect of the measurement system, including cables and modules~\cite{he2021channel}. First, the use of frequency multipliers in THz modules releases low-frequency cables from propagating THz signals. Therefore, the system is inherently free of the error caused by frequency mismatch of signals carried by cables. In addition to this, the measured the S21 parameter $S_{\rm measure}$, still contains the response of system modules, as
\begin{equation}\label{eq:S_measure}
    S_{\rm measure} = H_{\rm system} H_{\rm channel} G_{\rm AT},
\end{equation}
where $H_{\rm system}$ denotes the response of the measurement system and $H_{\rm channel}$ is the transfer function of the wireless channel. $G_{\rm AT}$ represent the cascaded antenna gain at Tx and Rx.
To take out the channel transfer function, we directly concatenated two RF fronts and measured the S21 parameter as calibration. To protect the device from high signal power, the antennas are removed and a 40-dB attenuator is added. Therefore, the S21 parameter given by the calibration, $S_{\rm calib}$, is composed of
\begin{equation}\label{eq:S_calib}
    S_{\rm calib} = H_{\rm system} G_{\rm attenuator},
\end{equation}
where $G_{\rm attenuator}$ is the response of the attenuator.
Combination of two equations \eqref{eq:S_measure} and \eqref{eq:S_calib} yields the channel transfer function, as
\begin{equation}
    H_{\rm channel} = \frac{S_{\rm measure}} {S_{\rm calib}} \times \frac{G_{\rm attenuator}}{G_{\rm AT}},
\end{equation}
where the antenna gain $G_{\rm AT}$ and the attenuator gain $G_{\rm attenuator}$ are known.

For each Tx-Rx position pair, the Rx scans the space at 36$\times$5 different directions. At each direction, we obtain 6001 samples of the channel transfer function (CTF) $H_{\rm channel}$. By inverse discrete Fourier transform (IDFT) of the CTF, 6001 samples of the channel impulse response (CIR) $h_{\rm channel}$ are derived. Each sample is associated with azimuth and elevation angles-of-arrival (AoA $\varphi$, EoA $\theta$) given by the direction of the Rx, and the time-of-arrival (ToA $\tau$) which is equal to the delay of the sample in the CIR.
We regard each sample as one possible MPC in the temporal domain and eliminate noise samples whose  power\footnote{Since the transfer function is derived after calibration, i.e., the system response is removed from the measured S21, the received power $P$, in essence, is the path loss (PL) with the unit of dB. For consistency, we denote $P$ as the power with the unit of dBm and regard the transmit power as 0~dBm. The actual transmit power is -10~dBm in this measurement campaign.} are lower than the threshold, given by
\begin{equation}
    P_{\rm TH} \text{ [dBm]} = \max\{P_{\rm m}-30, \text{NF}+10\},
\end{equation}
where $P_{\rm m}$ denotes the maximum power among measured samples at each position. NF is the average noise floor derived in CIR, which is -165~dBm in the long corridor (the LoS case) and -180~dBm in the wide corridor. The dynamic range, defined as the power difference between the largest MPC and the average noise floor, is about 65~dB. Despite the large dynamic range of the measurement system, we focus on MPCs within the dynamic range of 30~dB, in order to be compatible with the low dynamic range of general THz devices~\cite{han2022terahertz}.



\section{Channel Analysis and Characterization} \label{section: results}

In this section, the multi-path propagation in two corridors of the L-shaped hallway is traced and analyzed meticulously. Furthermore, channel characteristics, including path loss (PL), delay spread (DS) and azimuth spread of angle (ASA), and elevation spread of angle (ESA), at two frequency bands are analyzed and compared in depth. The equivalent alpha-beata path loss model for the NLoS hallway is proposed.

\subsection{Multi-path Propagation Analysis} \label{section: results_mpc}
\begin{figure}
    \centering
    \includegraphics[width=0.95\linewidth]{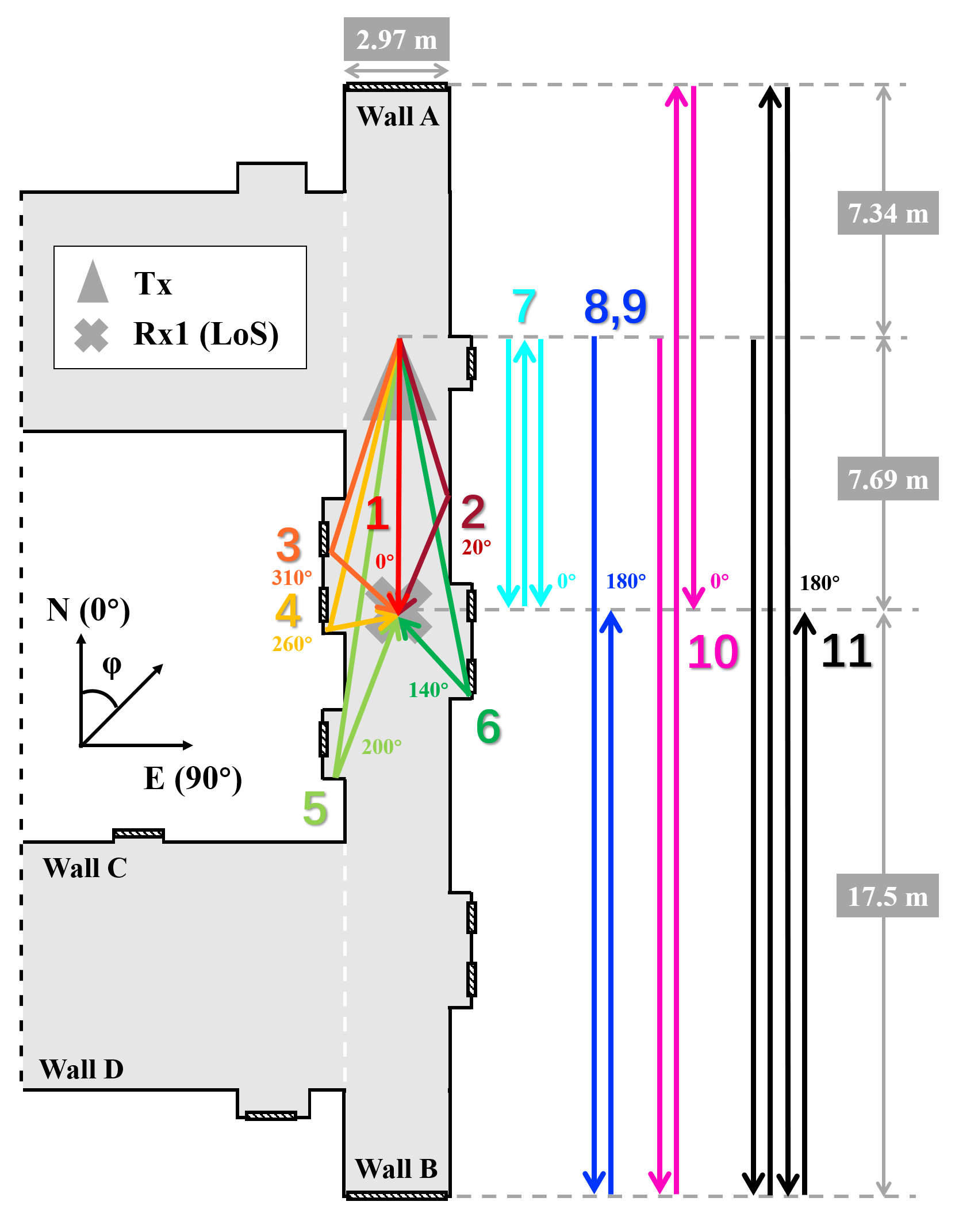}
    \caption{Major multi-path components at Rx1 in the indoor hallway scenario.}
    \label{fig:clustering_pt1}
\end{figure}

We analyze the multi-path propagation in the L-shaped hallway based on power-delay-angle profiles (PDAPs) and clustering results.
\rrev{The spatial and temporal distributions of major MPCs in the LoS hallway, taking Rx1 as an example, are first analyzed. Parameters of these MPCs are summarized in Table~\ref{tab:clustering_pt1} and illustrated in Fig.~\ref{fig:clustering_pt1}. Compared with the preliminary result in the short version~\cite{wang2022thz}, extra MPCs are recognized as follows. MPC~2, which travels 8.4~m, is reflected from the wall. MPC~3, 4, 5, and 6 are reflected from the indented offices distributed along the corridor. The values of the azimuth AoA and the ToA are in good accordance with the office locations in the campaign. Moreover, the longest path, MPC~11 arrives from Wall~B at 364~ns, which travels about 109~m. The MPC is reflected three times between Wall~A and Wall~B. The power of the MPC is -144.8~dB, on account of the strong reflection by Wall~A and Wall~B, which are actually metal doors.}

Next, we focus on the evolution of power distribution in the spatial domain in the NLoS hallway, and relevant observations motivate the modeling of major MPCs hybrid channel modeling in Section~\ref{section: hybrid_modeling}.

\begin{table}
    \centering
    \caption{Parameters of major multi-path components at Rx1.}
    \includegraphics[width=0.95\linewidth]{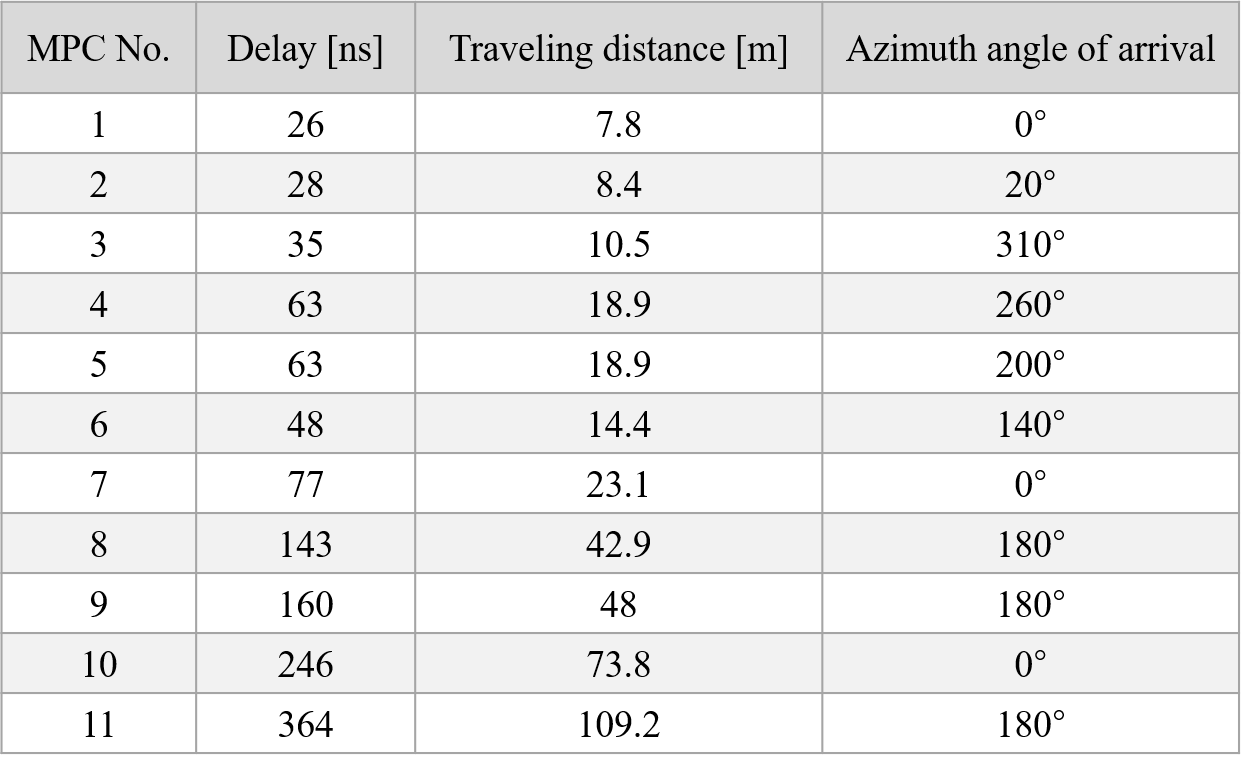}
    \label{tab:clustering_pt1}
\end{table}

\begin{figure}
    \centering
    \includegraphics[width=\linewidth]{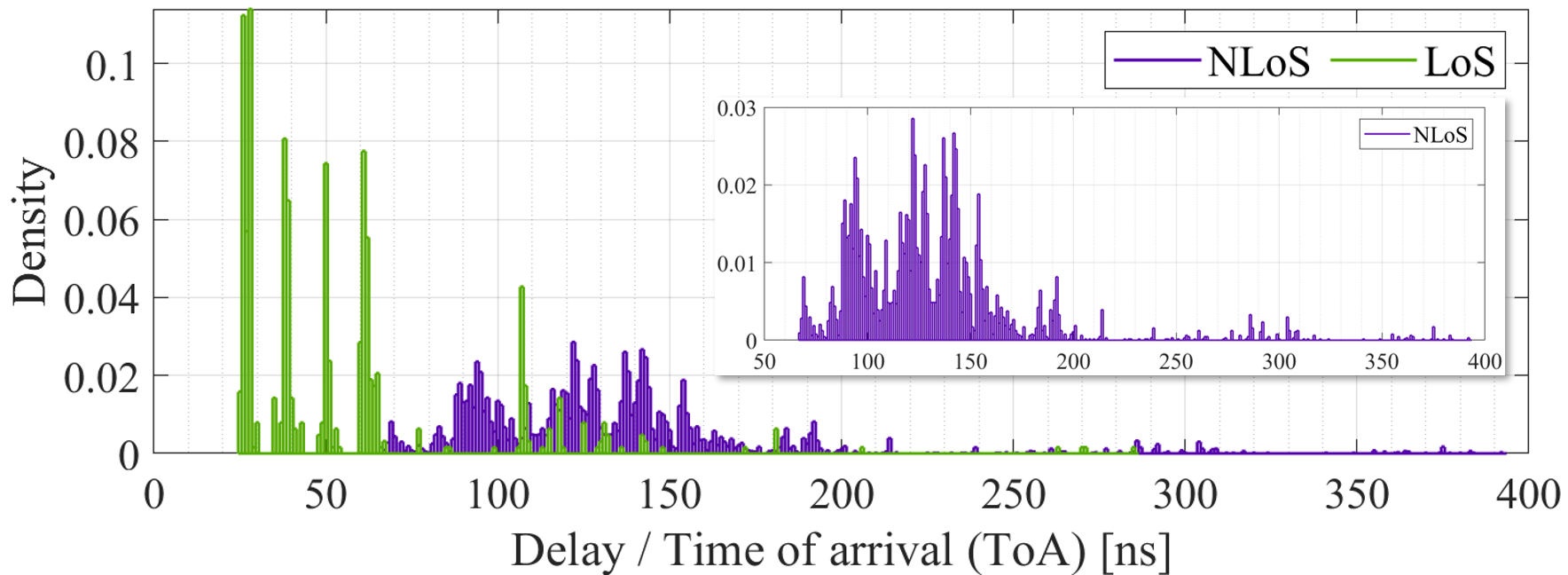}
    \caption{Probability density of MPC ToA in LoS and NLoS cases.}
    \label{fig:delay_distribution}
\end{figure}
\begin{figure}
    \centering
    \subfigure[Far-NLoS, the first row.]{
    \includegraphics[width=0.45\linewidth]{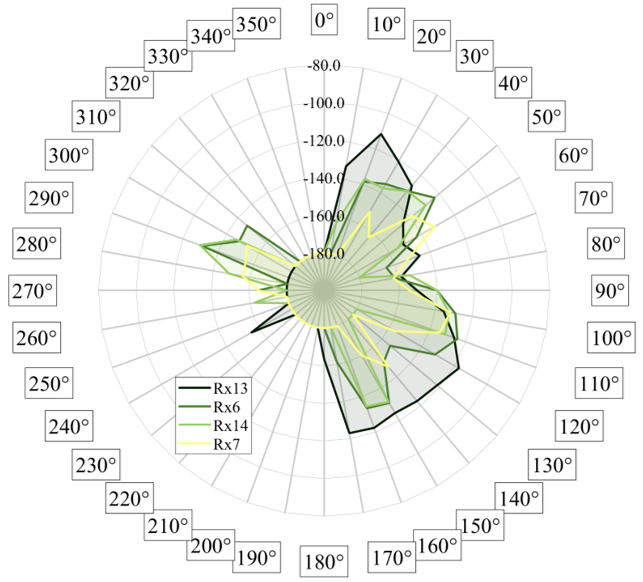}
    }
    \subfigure[Near-NLoS, the first row.]{
    \includegraphics[width=0.45\linewidth]{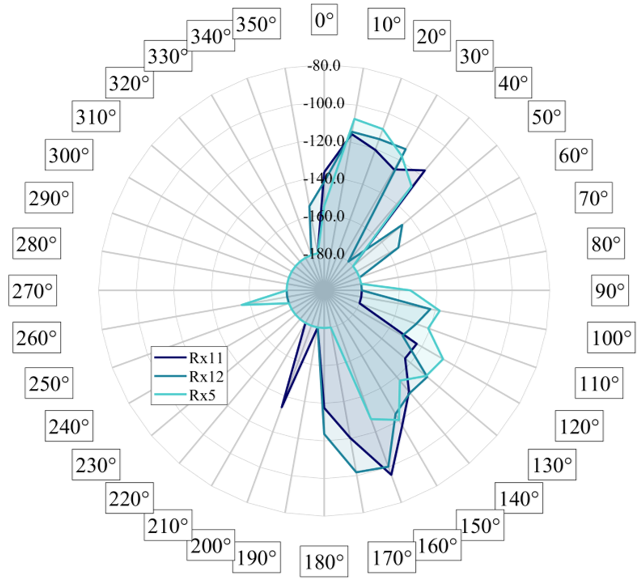}
    }
    \\
    \subfigure[Far-NLoS, the second row.]{
    \includegraphics[width=0.45\linewidth]{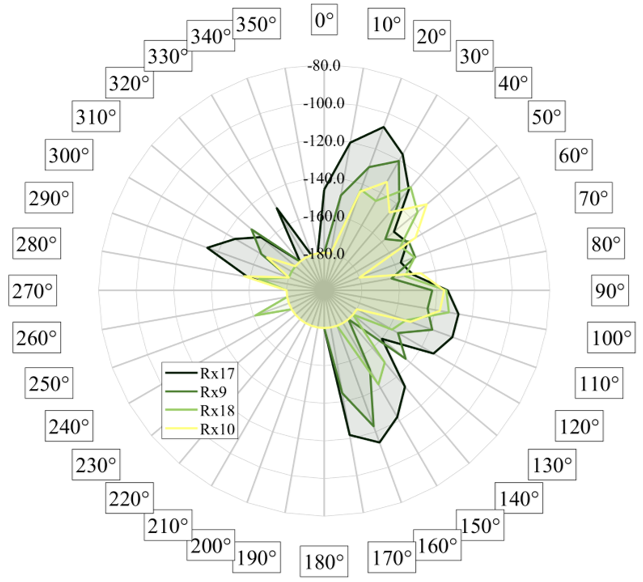}
    }
    \subfigure[Near-NLoS, the second row.]{
    \includegraphics[width=0.45\linewidth]{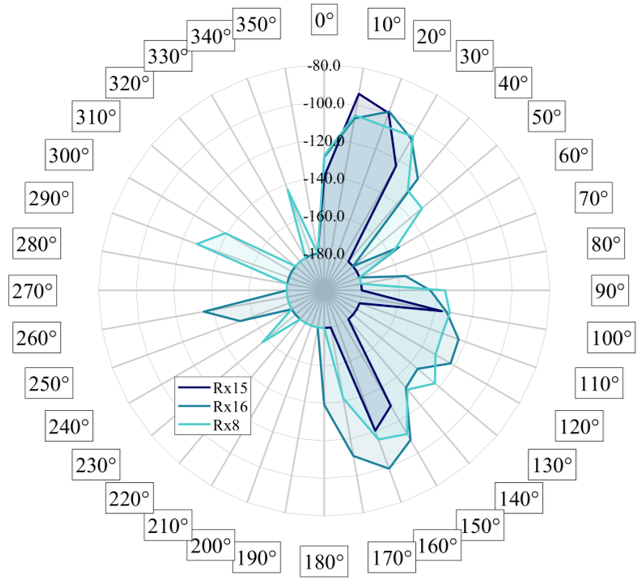}
    }
    \caption{Sum of power [dBm] at different azimuth angles of arrival at NLoS Rx in the indoor hallway scenario.}
    \label{fig:p_sum_nlos_indoor}
\end{figure}
\begin{figure}
    \centering
    \includegraphics[width=0.8\linewidth]{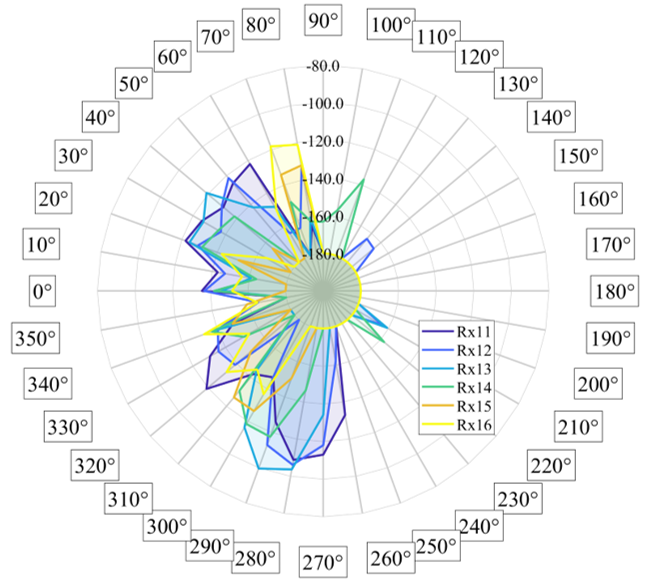}
    \caption{Sum of power [dBm] at different azimuth angles of arrival at NLoS Rx in the outdoor street scenario.}
    \label{fig:p_sum_nlos_outdoor}
\end{figure}

The probability density of MPCs' ToA in LoS and NLoS cases are shown in Fig.~\ref{fig:delay_distribution}, respectively.
First, in the NLoS case, MPCs mainly come from the corner of two corridors and travel at least 19~m to reach the near-NLoS points next to the corner. Besides, NLoS MPCs that travel a long distance (more than 60~m in our case) are typically too weak to be detected. Therefore, the majority of MPCs arrive at Rx with the ToA ranging from 65~ns to 200~ns.
Furthermore, due to the absence of the dominant LoS ray, the scattering effect becomes more distinct in the NLoS case. This phenomenon is verified by the flattening of ToA density, i.e. the spread of ToA, in the NLoS case as shown in Fig.~\ref{fig:delay_distribution}, compared with ToA values concentrated at deterministic values in the LoS case.

Then, we analyze the spatial distributions of MPCs in the wide corridor. In Fig.~\ref{fig:p_sum_nlos_indoor}, the power of MPCs with different ToAs is summed up at each azimuth AoA and NLoS Rx position. The four subfigures correspond to the layout of NLoS Rx positions. The observations are summarized as follows. First, the majority of MPCs come from the direction of the corner, i.e., the azimuth AoA ($\varphi$) are concentrated within the range from 0 to 180$^\circ$. In specific, as the Rx position moves farther away from the corner, the values of azimuth AoA tend to converge towards the east with $\varphi=90^\circ$, i.e., the center line of the wide corridor. Second, we can observe the phenomenon of beam-like power distribution in Fig.~\ref{fig:p_sum_nlos_indoor}. Each beam can also be regarded as a cluster with respect to AoA, and the beam center denotes the direction of the dominant ray received at the Rx. Third, when the Rx moves away from the corner, the beams first become wider in the near-NLoS case as Fig.~\ref{fig:p_sum_nlos_indoor}(b)(d) and then narrower in the far-NLoS case as Fig.~\ref{fig:p_sum_nlos_indoor}(a)(c). The phenomenon indicates that the power becomes more spread in the angular domain inside near-NLoS range and then concentrates when the Rx goes into the far-NLoS range. The power spread increases as scattered rays become relatively more significant, while the trend is reversed after the scattered rays are drowned below the noise floor.


Interestingly, a similar phenomenon is observed in outdoor street measurement results. Channel measurement and multi-path propagation analysis in the outdoor street are discussed in~\cite{wang2022terahertz}. In this scenario, the corner (junction) of two streets is at the west with azimuth AoA $\varphi=0^\circ$. As shown in Fig.~\ref{fig:p_sum_nlos_outdoor}, in the NLoS case, as the Rx position moves farther away from the corner, from Rx11 to Rx16, the values of azimuth AoA tend to converge towards the direction of the corner, i.e., $\varphi=0^\circ$. Moreover, the beam-like power distribution also exists and the beams become narrower.

\begin{figure}
    \centering
    \includegraphics[width=\linewidth]{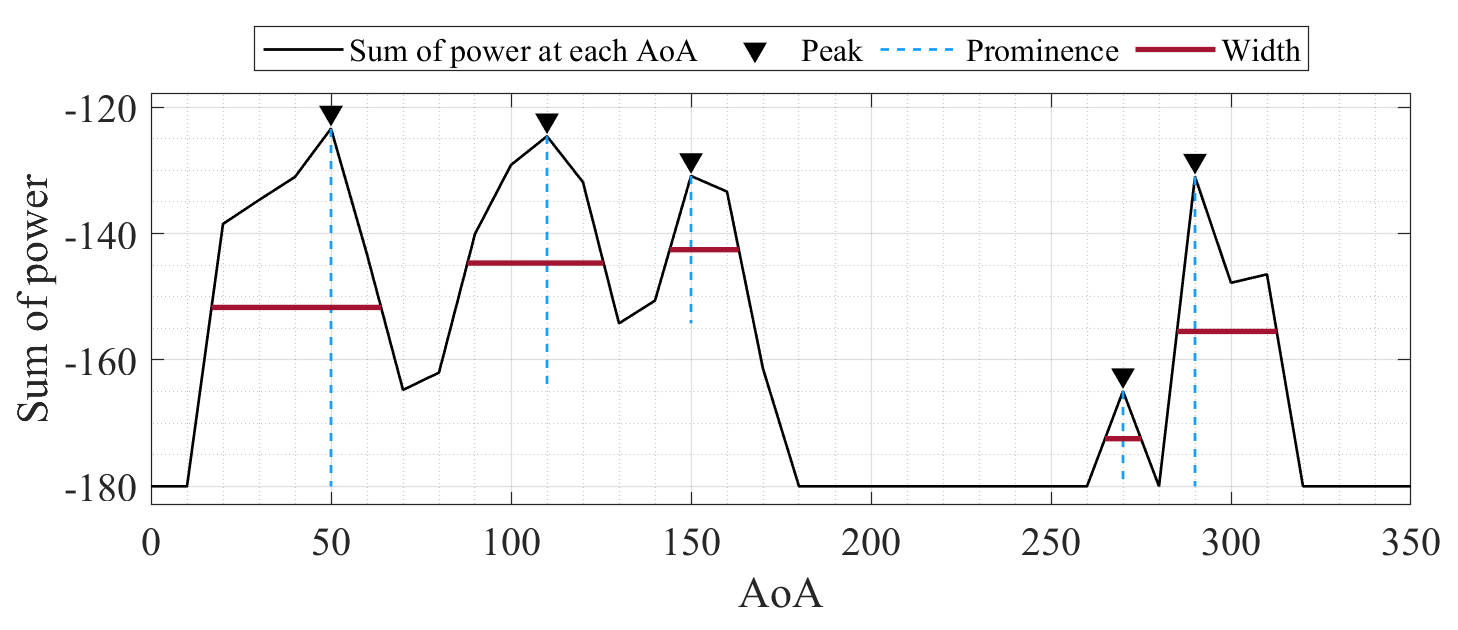}
    \caption{Peaks at Rx6 at 306-321~GHz.}
    \label{fig:findpeak_sample}
\end{figure}

\begin{table*}
    \centering
    \caption{Summary of beam parameters in indoor hallway and outdoor street at two frequency bands.}
    \includegraphics[width=\linewidth]{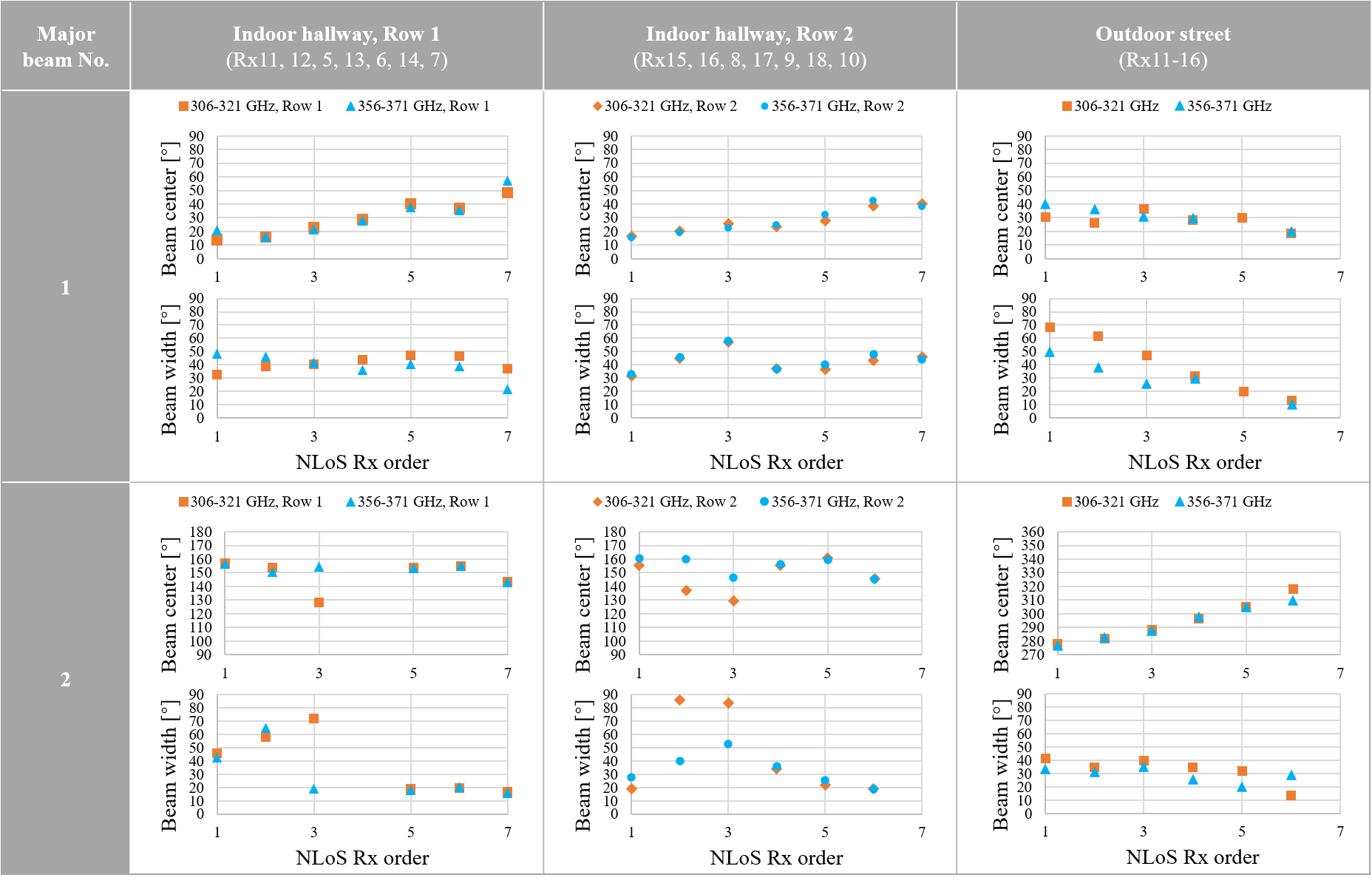}
    \label{fig:beam_summary}
\end{table*}

\rev{The phenomenon is characterized by finding the peaks in the sum of the power at each AoA. In order to tolerate a small degree of fluctuation, the peaks are recognized as beams when their prominence is larger than 15~dB. The measure of prominence, different from the height, indicates how the peak stands out compared with surrounding peaks. The algorithm is described as follows. Once we identify a candidate peak at $\varphi_{\rm lm}$ with a local maximum $P_{\rm lm}$, we search for the nearest AoAs where the power is at the same level $P_{\rm lm}$ from two directions, i.e., $\varphi_{\rm left}$ and $\varphi_{\rm right}$. As the value of AoA is cyclic, we stop when the search passes through 180$^\circ$, instead of stopping at the endpoints of 0$^\circ$ and 350$^\circ$. The two AoAs define borders of two intervals near the location of the local maximum, and we examine the minimum of the power in each interval, respectively. The higher of the two minima defines the reference level $P_{\rm ref}$, and the difference between $P_{\rm lm}$ and $P_{\rm ref}$ is the prominence. As mentioned, the candidate peak with a prominence larger than 15~dB is recognized as a beam. The beam width is equal to the distance between AoAs where the power sum descends from the peak by half of the prominence. The beam center is defined as the center of the two demarcation points.}

The result at Rx6 at 306-321~GHz is shown in Fig.~\ref{fig:findpeak_sample}.  Parameters of major beams in indoor hallway and outdoor street at two frequency bands are summarized in Table~\ref{fig:beam_summary}. Beam centers are regarded as the AoA of dominant MPCs in hybrid modeling in Sec.~\ref{section: hybrid_modeling}, while beam widths characterize the spatial spread of power. The NLoS Rx are ordered from the close-by Rx to those far off.
We observe that first, the trends of beam parameters, both the center and the width, are consistent for each row of NLoS Rx in either scenario. Second, beam centers at different Rx positions converge to the direction of the corner in each scenario. Third, in the indoor hallway where the power is larger than the counterpart in the outdoor street, as the NLoS Rx moves away from the corner, the width first increase since more rays become within the 30-dB dynamic range when the maximum power decrease. After the maximum power become small so that the 30-dB dynamic range includes the majority of scattering rays, as the case in the outdoor street, the width decreases as the NLoS Rx moves farther away from the corner.

\subsection{Characteristic Analysis}

\subsubsection{Path loss}

The path loss is divided into best direction path loss and omni-directional path losses. For each pair of Tx and Rx positions, the best direction path loss is defined as the loss of the MPC from the best direction which has the strongest received power. By contrast, the omni-directional path loss sums the received power from all directions. The two path losses are respectively given by
\begin{subequations}
    \begin{equation}
    \text{PL}_{\rm best} \text{ [dB]} = -10\cdot\log_{10} \left( \max_{i,j}\{P_{i,j}\} \right),
    \end{equation}
    \begin{equation}
    \text{PL}_{\rm omni} \text{ [dB]} = -10\cdot\log_{10} \left( \sum_{i,j}P_{i,j} \right),
    \end{equation}
\end{subequations}
where $P_{i,j}$ denotes the received power from the $i^{\rm th}$ azimuth angle and the $j^{\rm th}$ elevation angle, as
\begin{equation}\label{eq:power_sum}
    P_{i,j} = \sum_{t}|h_{i,j,t}|^2,
\end{equation}
where $h_{i,j,t}$ represents the CIR at the $t^{\rm th}$ sample time point from, the $i^{\rm th}$ azimuth and the $j^{\rm th}$ elevation angles.

The close-in free space reference distance (CI) and $\alpha$-$\beta$ path loss models are invoked based on the results of best direction and omni-directional path losses, respectively. The two path loss models are expressed as
\begin{subequations}
\begin{align}
&\text{PL}^{\rm CI} = 10\times\text{PLE}\times\log_{10}\frac{d}{d_0}+\text{FSPL}(d_0,f)+X^{\rm CI}_{\sigma_{\rm SF}},\\
&\text{PL}^{\alpha\beta} = 10\times\alpha\times\log_{10}d+\beta+X^{\rm \alpha\beta}_{\sigma_{\rm SF}},
\end{align}
\end{subequations}
where PLE is the path loss exponent, $\alpha$ denotes the slope coefficient, and $\beta$ represents the optimized path loss offset in dB. $d$ denotes the distance between Tx and Rx, and $d_0$, which is 1~m in this work, represents the reference distance. $X_{\sigma_{\rm SF}}$ is a zero-mean Gaussian random variable with standard deviation $\sigma_{\rm SF}$ in dB, indicating the fluctuation caused by shadow fading. Furthermore, the free-space path loss (FSPL) in dB is given by the Friis’ law, as
\begin{equation}
    \text{FSPL}(d,f) = -20\times\log_{10}\frac{c}{4\pi f d},
\end{equation}
where $f$ denotes the frequency, and $c$ is the speed of light.
We determine the value of PLE in the CI model and values of $\alpha$ and $\beta$ in the $\alpha$-$\beta$ model by minimizing $\sigma_{\rm SF}$, respectively.

\begin{figure*}
    \centering
    \subfigure[306-321~GHz.]{
    \includegraphics[width=0.8\linewidth]{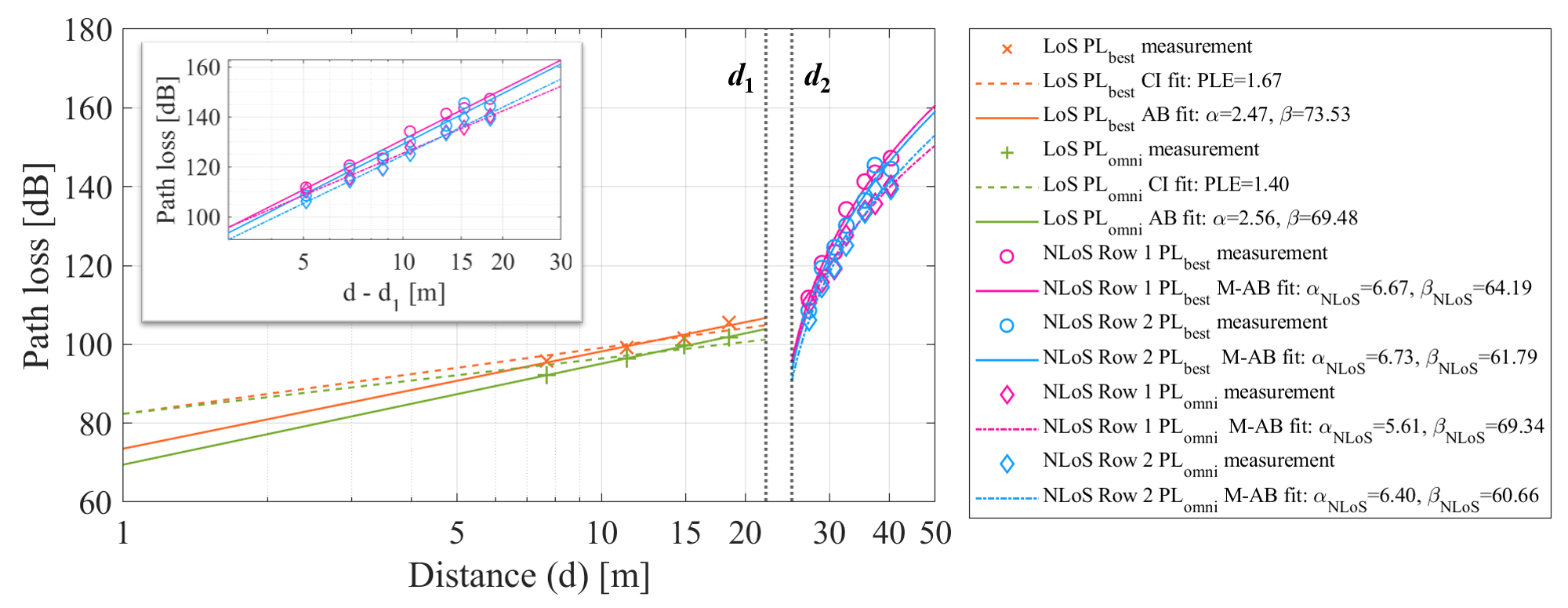}
    }
    \\
    \subfigure[356-371~GHz.]{
    \includegraphics[width=0.8\linewidth]{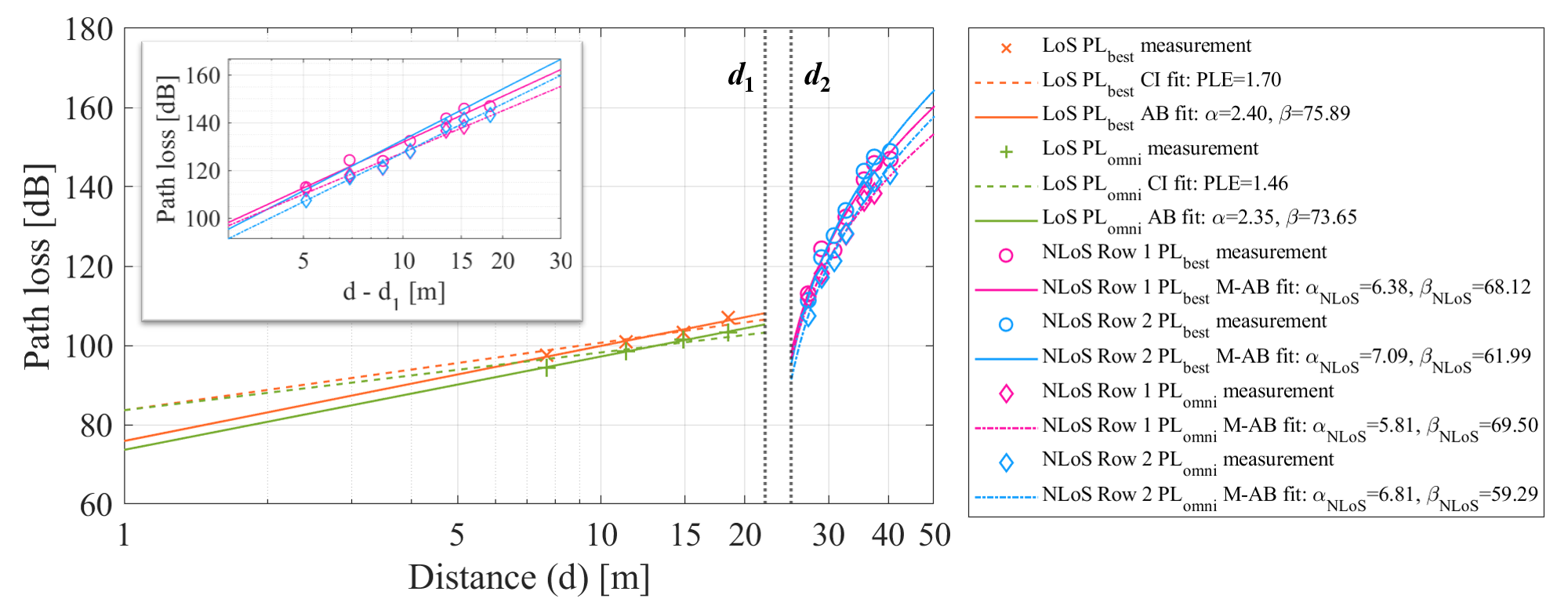}
    }
    \caption{\rrev{Path loss measurement and fitted CI, $\alpha$-$\beta$, and modified $\alpha$-$\beta$ (M-AB) path loss models in the indoor hallway scenario.}}
    \label{fig:pl_indoor_combine}
\end{figure*}
\begin{figure*}
    \centering
    \subfigure[306-321~GHz.]{
    \includegraphics[width=0.8\linewidth]{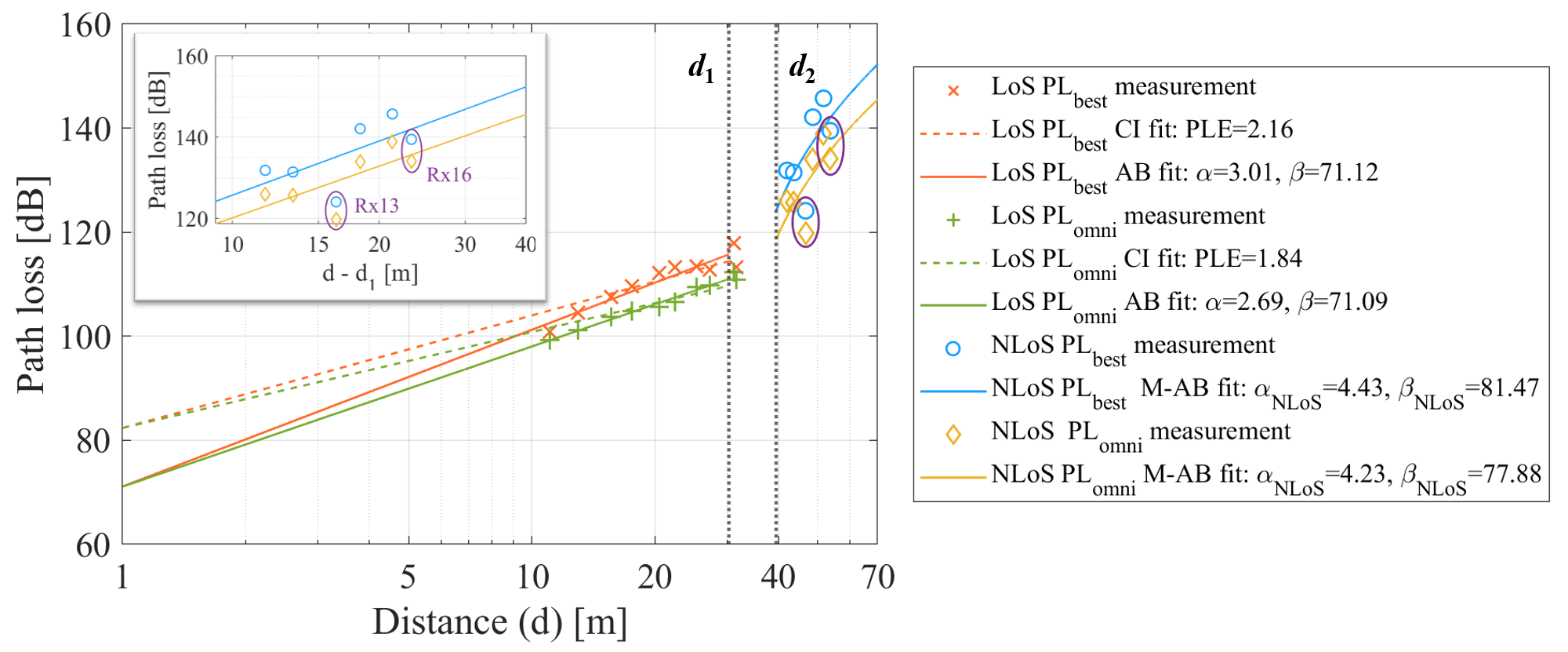}
    }
    \\
    \subfigure[356-371~GHz.]{
    \includegraphics[width=0.8\linewidth]{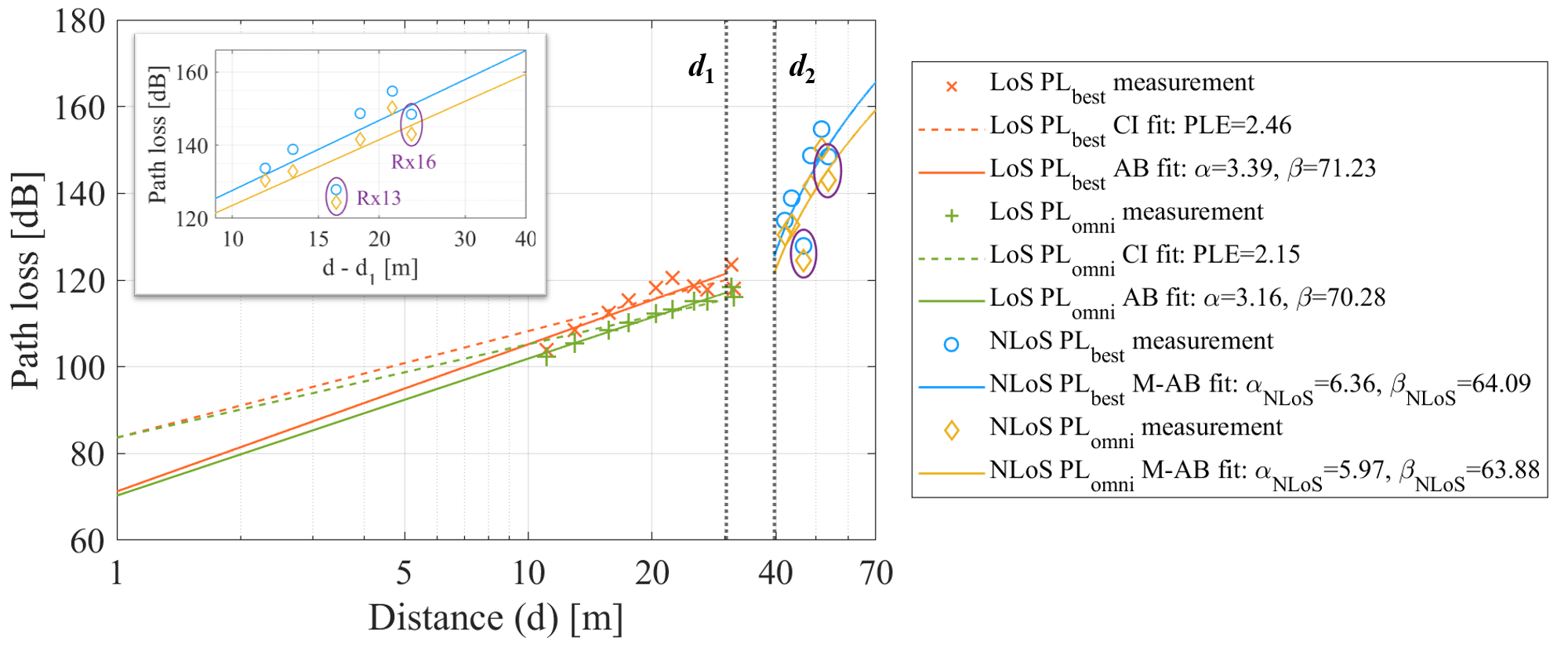}
    }
    \caption{\rrev{Path loss measurement and fitted CI, $\alpha$-$\beta$, and modified $\alpha$-$\beta$ (M-AB) path loss models in the outdoor street scenario.}}
    \label{fig:pl_outdoor_combine}
\end{figure*}

Measurement and fitting results of path losses in the LoS case are shown in Fig.~\ref{fig:pl_indoor_combine} and summarized in Table~\ref{tab:characteristic_noncluster}. \rrev{Two vertical lines correspond to $d_{1}=22.09$~m and $d_{2}=25.06$~m in the L-shaped indoor scenario.}
The CI model fitting results yield that at 306-321~GHz, the PLE value is 1.67 for the best direction path loss and 1.40 for the omni-direction path loss, respectively. At 356-371~GHz, the values increase to 1.70 and 1.46, correspondingly.
Justifications are elaborated as follows.
First, due to the waveguide effect, besides the LoS ray, strong reflected paths with different ToAs can also be detected from the best direction. Thus, the received power is increased by the summation of received power in the \textit{time} domain, as expressed in~\eqref{eq:power_sum}. As a result, the best direction path loss is reduced, and the PLE value is smaller than 2, which is the PLE for ideal free-space path loss.
Second, the omni-directional received power sums the received power in both \textit{time} and \textit{angular} domains. To be specific, besides paths from the best direction, reflected paths from the south, and scattered paths from all other directions are included in the summation of~\eqref{eq:power_sum}, resulting in a smaller PLE for the omni-directional path loss.
Furthermore, according to the $\alpha$-$\beta$ model fitting results, optimal path loss offsets $\beta$ are larger than the reference free space path loss at 1~m, and slope coefficients $\alpha$ are all larger than 2.

As depicted in Fig.~\ref{fig:pl_indoor_combine}, in the NLoS case, the values of Tx-Rx distances $d$ are relatively concentrated near 20~m. Therefore, despite the abundance of NLoS Rx points, CI and $\alpha$-$\beta$ path loss models that depend on the Tx-Rx distance are not suitable for the NLoS case in L-shaped indoor hallway and outdoor street scenarios.
Based on the observation in the indoor hallway in Fig.~\ref{fig:p_sum_nlos_indoor} and Section~\ref{sec:MPC_analysis_NLoS}, since the majority of power comes from the east corner, we assume a virtual source at the east end of the wide corridor and propose the modified $\alpha$-$\beta$ (M-AB) path loss model for the NLoS corridor \rrev{in terms of $d-d_{1}$}, which is expressed as
\begin{equation}
    \text{PL}^{\alpha\beta}_{\rm NLoS} = 10\times\alpha_{\rm NLoS}\times\log_{10}(d-d_{1})+\beta_{\rm NLoS}+X^{\rm NLoS}_{\sigma_{\rm SF}},
\end{equation}
where $\alpha_{\rm NLoS}$ and $\beta_{\rm NLoS}$ are the equivalent exponent and best-fit path loss offset. \rrev{As illustrated in Fig.~\ref{fig:deployment}(a), $d-d_{1}$ is the distance of the Rx to the east end of the wide corridor. According to the deployment, NLoS Rx positions are separated into two rows, and for each row, the values of $d-d_{1}$ are 5.09~m, 6.89~m, 8.69~m, 10.49~m, 13.49~m, 15.29~m, and 18.30~m in the ascending order.}

The fitting results for NLoS Rx in two rows in the indoor hallway are shown in Fig.~\ref{fig:pl_indoor_combine}. At both frequency bands, the modified $\alpha$-$\beta$ path loss model \rrev{with respect to $d-d_{1}$} is a good fit for the NLoS hallway.
The equivalent exponent $\alpha_{\rm NLoS}$ in the indoor NLoS hallway is about 6-7 for best direction path loss, smaller than 7-8 for omni-directional path loss. The equivalent optimal offset $\beta_{\rm NLoS}$ is a case-by-case parameter depending on the location of Tx and the dimension of the long corridor where the Tx is located, \rrev{i.e., $d_1$ and $d_2$}. In this scenario, the value of the equivalent optimal offset $\beta_{\rm NLoS}$ is around 60-70~dB.
Furthermore, limited by the amount of measurement data, the proposed model simulates the relation between the equivalent distance and the path loss in each row, which does not explore the effect on path loss between different rows.
\rrev{So far, no distinctive difference is observed between the two rows. Preliminary results on path loss difference between each pair of two aligned Rx in Row~1 and Row~2 are given as follows.} Based on the measurement result in this work, at 306-321~GHz, the average difference of path losses between aligned Rx position pairs in two rows is 2.76~dB for the best direction path loss and 1.92~dB for the omni-directional counterpart. At 356-371~GHz, these values decrease to 2.14~dB and 1.58~dB, correspondingly.

To generalize the proposed modified $\alpha-\beta$ path loss model, we apply it in the outdoor street~\cite{wang2022terahertz}. \rrev{In the measurement deployment of the outdoor street, $d_{1}$ is 30.36~m, $d_{2}$ is 39.59~m and the values of $d-d_{1}$ for NLoS Rx are} 11.67~m, 13.30~m, 16.31~m, 18.27~m, 21.27~m, and 23.27~m in the ascending order. The fitting results for NLoS Rx in the outdoor street are shown in Fig.~\ref{fig:pl_outdoor_combine}. The modified $\alpha$-$\beta$ path loss model \rrev{with respect to $d-d_{1}$} is also a good fit for the NLoS case in the outdoor street scenario, analogy to the counterpart in the indoor hallway scenario. In particular, path losses at Rx13 and Rx16 are smaller and deviate from the linear fit for the following reason. First, at Rx13, extra transmission rays through the vehicle are detected. Second, at Rx16, the reflection ray from the northern building is enhanced, compared with the counterparts at other Rx positions. These observations are already illustrated in~\cite{wang2022terahertz}, and result in the reduction of path losses at the two points.
The equivalent exponent $\alpha_{\rm NLoS}$ in the outdoor NLoS street is about 4 at 306-321~GHz, smaller than around 6 at 356-371~GHz. The equivalent optimal offset $\beta_{\rm NLoS}$ in this scenario is around 64~dB at 306-321~GHz and about 80~dB at 356-371~GHz.
In both sub-bands, compared with the counterparts in the indoor L-shaped scenario, the equivalent optimal offset $\beta_{\rm NLoS}$ in the outdoor L-shaped scenario is larger and the equivalent exponent $\alpha_{\rm NLoS}$ is smaller.

\begin{table*}
    \centering
    \caption{Summary of channel characteristics.}
    \subfigure[Channel characteristics at 306-321~GHz.]{
    \includegraphics[width=0.9\linewidth]{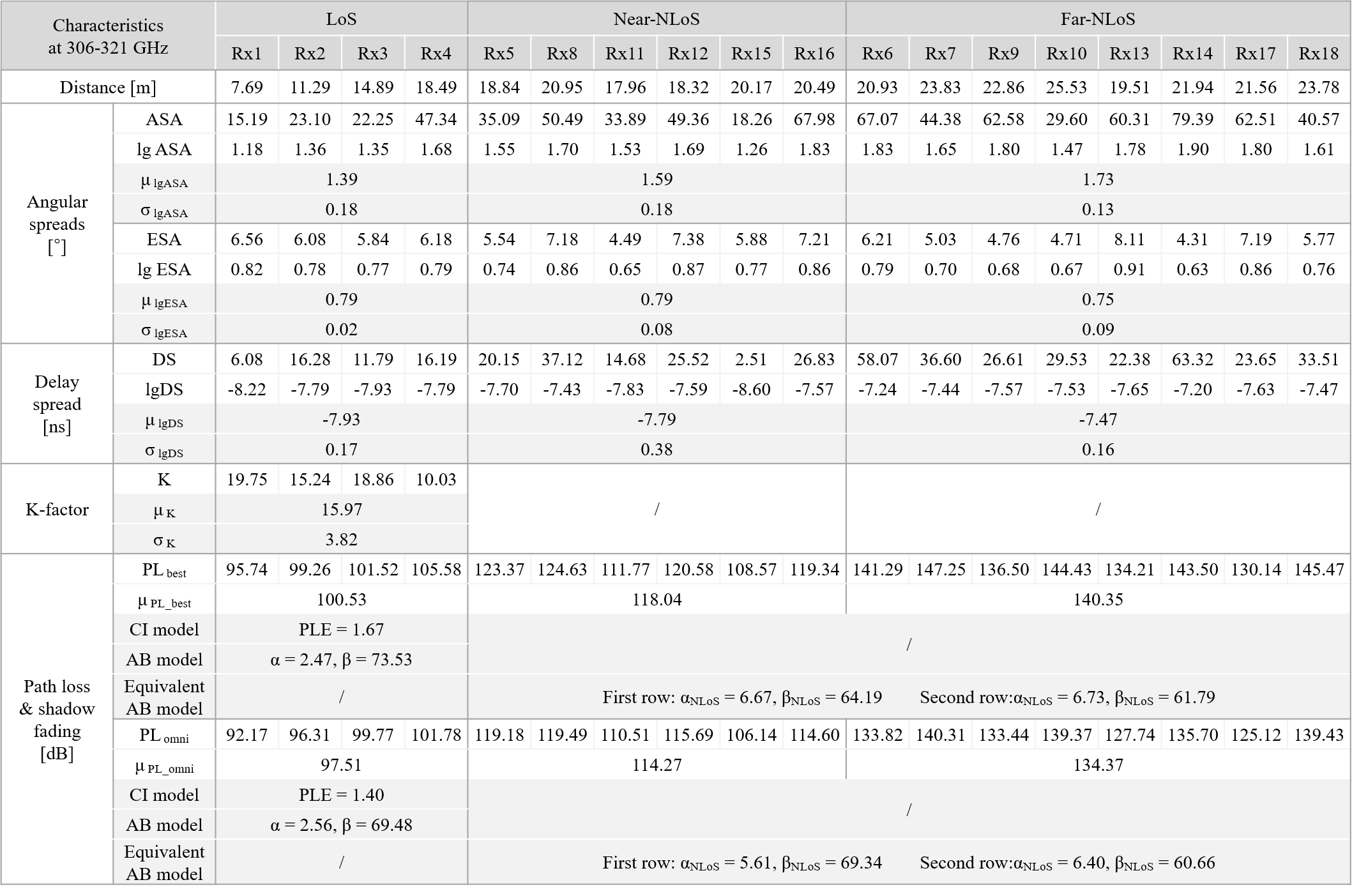}
    }
    \\
    \subfigure[Channel characteristics at 356-371~GHz.]{
    \includegraphics[width=0.9\linewidth]{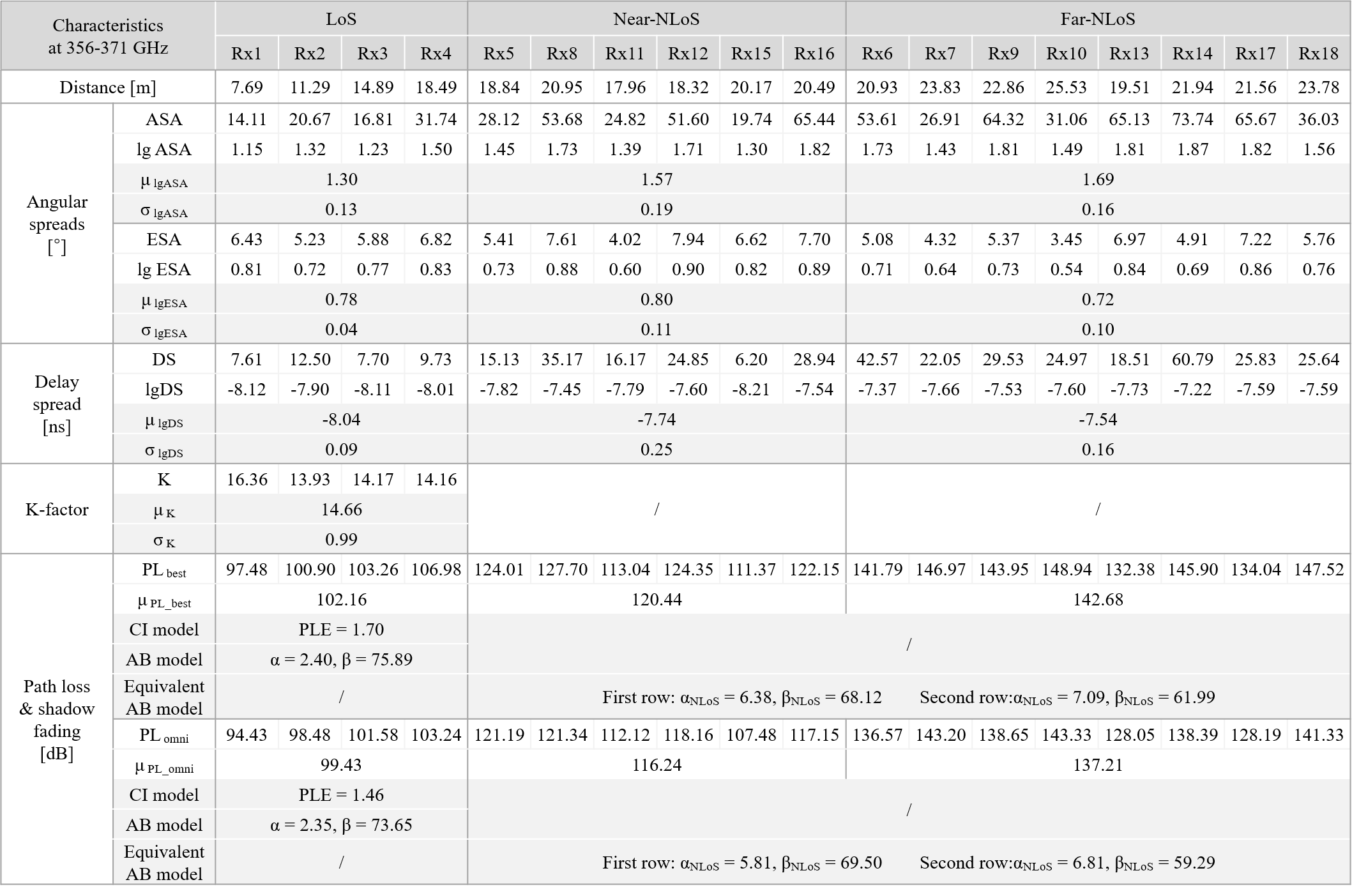}
    }
    \label{tab:characteristic_noncluster}
\end{table*}

\subsubsection{Delay and angular spreads}
We use the root-mean-square (RMS) delay spread and angular spreads to measure the power dispersion of MPCs in temporal and spatial domains, respectively. The results are summarized in Table~\ref{tab:characteristic_noncluster}.
In the LoS case, the values of DS are less than 20~ns. At Rx1, the power of the LoS ray is high. As a result, the reflection from the south end of the corridor is not within the 30-dB dynamic range, and DS at Rx1 is lower.
The value of ASA increases as the position of Rx moves closer to the corner, which is caused by MPCs which are reflected and scattered between the corner and the cart. These MPCs arrive at the Rx with the azimuth AoA around 180$^\circ$. As a result, the increase of received power from the south (with $\varphi$ near $180^\circ$) counterweights the power of MPCs from the north (with $\varphi$ near $0^\circ$ or $360^\circ$), and results in a larger ASA. The average DS and ASA in the LoS case are 12.58~ns and 26.97$^\circ$ at 306-321~GHz. At 356-371~GHz, average DS and ASA in the LoS case decrease to 9.39~ns and 20.84$^\circ$, respectively.

In NLoS cases, since most MPCs have power comparable to the noise floor, values of DS and ASA are highly dependent on the dominated MPCs. At near-NLoS Rx positions, best direction path losses are lower, indicating the existence of high-power reflected MPCs, which are dominant in the near-NLoS case.
When Rx moves farther away into far-NLoS positions, these MPCs become weaker and are overwhelmed by scattering MPCs between Wall~C and Wall~D.
Furthermore, some MPCs are buried in the noise, which renders inconclusive changes of DS and ASA. Though, in the general sense, the average DS and ASA are the largest in the far-NLoS case, followed by the near-NLoS case and then the LoS case. At 356-371~GHz, ASA and DS become smaller compared with the counterparts at 306-321~GHz, except for DS in the near-NLoS case which is comparable at two frequency bands.

\section{Hybrid Channel Modeling with the evolving model for the THz NLoS Hallway} \label{section: hybrid_modeling}


\rev{The methodologies for physical wireless channel modeling is categorized as deterministic, statistical, and hybrid approaches~\cite{han2022terahertz}. While deterministic channel modeling has high accuracy at the cost of high time and resource consumption, statistical channel modeling compromises accuracy for low complexity and general applicability. The trend of channel modeling towards 6G is to develop hybrid methods by combining two approaches.}

Deterministic simulation results can improve pure statistical channel modeling for the following reasons.
First, the software can simulate EM propagation and provides channel characteristics in indoor environments. The output includes the direction of departure, the direction of arrival, the time of arrival, and the received power, which accord with those derived from the measurement data. Therefore, it enables parallel or combined analysis of results from the measurement and the simulation. Moreover, the ray-tracing (RT) simulator provides the traveling path of each ray, including the number, position, and type of interactions, which has explicit physical meaning. This advantage over the measurement provides additional information in helping identify interaction objects (IOs) in the environment.

However, there is disagreement between simulation and measurement results, due to the following facts: (a) The floor plan is simplified; (b) the accurate antenna pattern is unavailable; (c) the interaction of MPCs, especially the scattering effect, is inaccurate owing to the uncertainty of precise interaction models and material parameters of interaction objects at the target frequency band~\cite{he2019design}; (d) the temporal and/or spatial resolution is different.
\rev{In mmWave and THz bands, the simulation accuracy of dominant MPCs is verified~\cite{zhang2022geometry,ding2020ray,zhang2023deterministic}.}

Still, deviation of angle, delay and power of other MPCs between simulation and measurement is also observed in these works. \rev{The discrepancy becomes worse in the THz band, where the material property is inadequate, and especially in the NLoS case, on account of the long propagation path and the inaccuracy of the scattering model.}
Therefore, we develop a deterministic-statistical hybrid channel modeling method for the THz indoor channel~\cite{chen2021channel}. \rev{Specifically, clusters that are associated with dominant MPCs are called RT clusters, where dominant MPCs, i.e., the LoS path and wall-reflection paths, can be captured by ray-tracing and regarded as cluster centers. Other paths in RT clusters and paths in non-RT clusters are generated in a statistical approach.}
In the following part, we discuss the matching degree between the simulation result and measurement data in the THz band, and suggest a hybrid channel model for the THz NLoS hallway scenario based on the previous observation in Section~\ref{section: results_mpc}.

\subsection{Deterministic Part based on Ray-Tracing Simulation}
\begin{figure}
    \centering
    \includegraphics[width=0.9\linewidth]{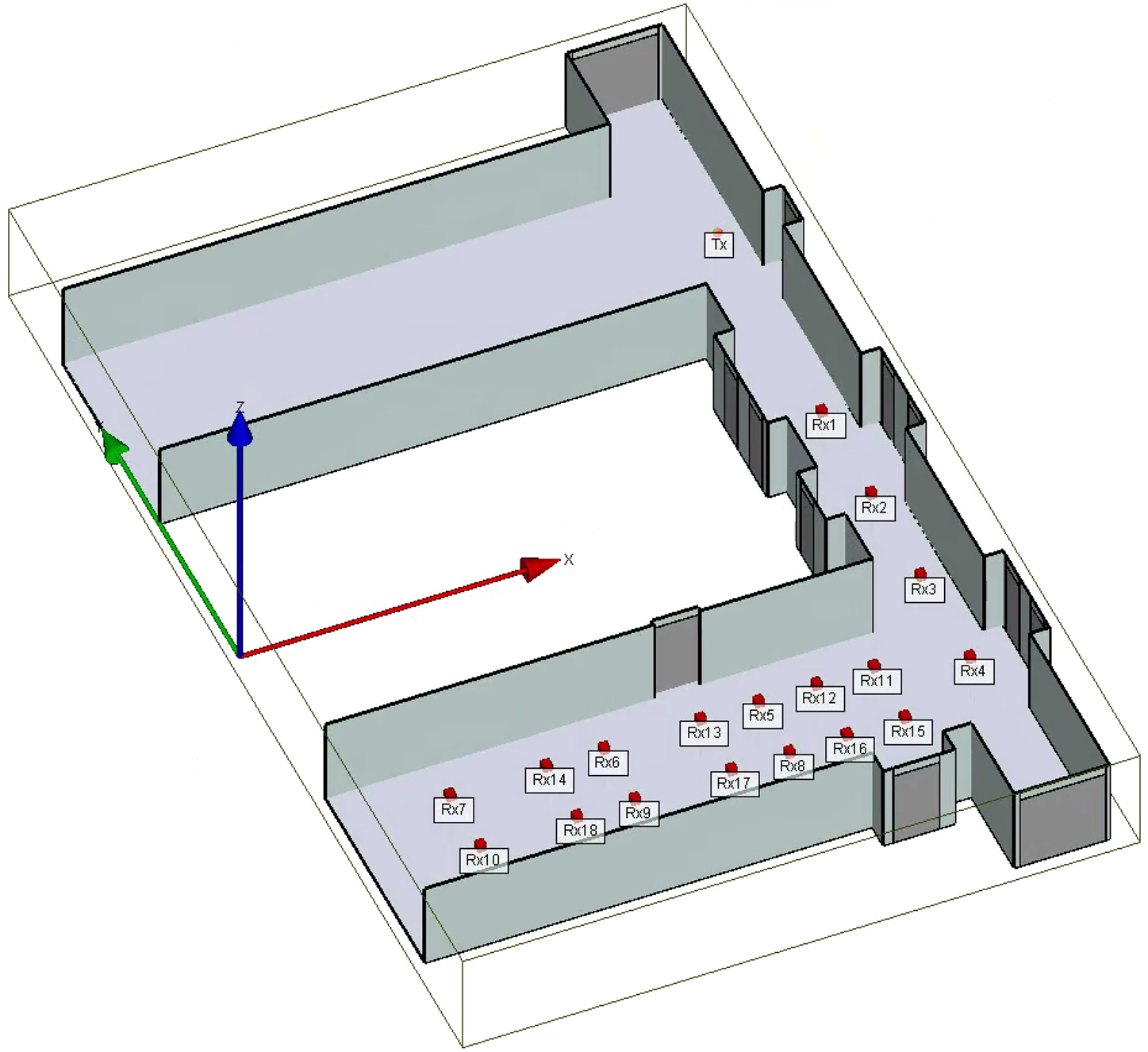}
    \caption{Simulation deployment.}
    \label{fig:sim}
\end{figure}

The simulation is implemented in Wireless InSite, with the deployment depicted in Fig.~\ref{fig:sim}.
\begin{figure}
    \centering
    \subfigure[Rx11.]{\includegraphics[width=\linewidth]{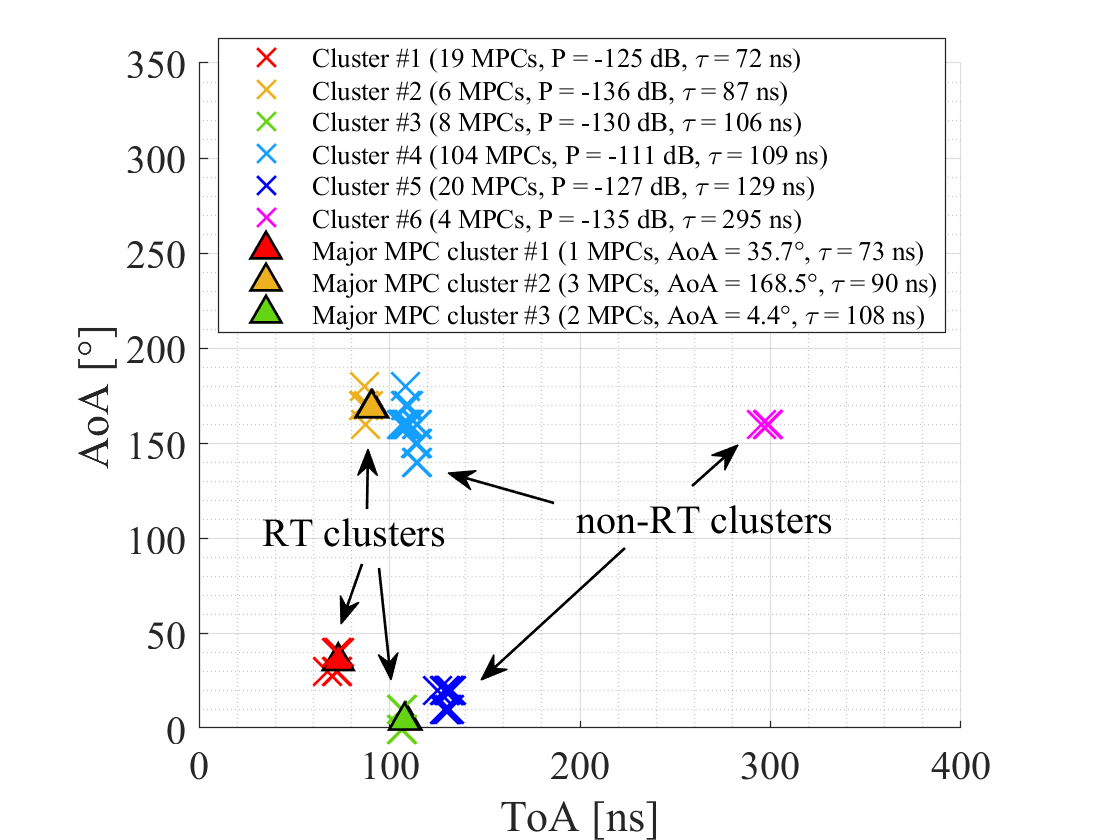}}
    \subfigure[Rx15.]{\includegraphics[width=\linewidth]{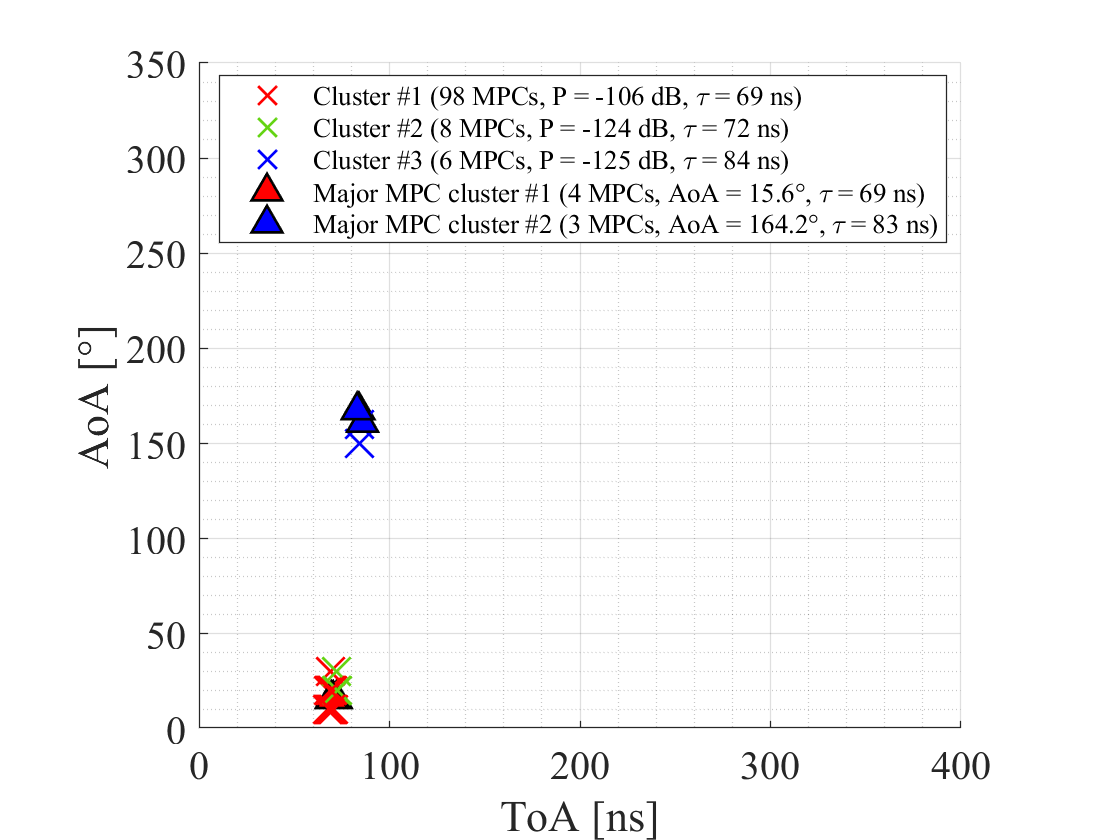}}
    \caption{Measurement result (cross mark) and simulated major MPCs (triangle mark) in the NLoS hallway at 306-321~GHz.}
    \label{fig:mea_rt_major}
\end{figure}
Utilizing the information about the interaction type along the traveling path given by the simulation result, major MPCs only include LoS and reflected paths. This accords with the fact that the LoS path and wall-reflection paths are dominant in the THz indoor channel~\cite{chen2021channel}. Moreover, this evades the inaccuracy of simulation of the scattering effect.

The measurement result and the simulation result of major MPCs at Rx11 and Rx15 (the first near-NLoS Rx position in each row in the NLoS hallway) at 306-321~GHz are shown in Fig.~\ref{fig:mea_rt_major}. The result verifies the accuracy of simulated multi-bounce reflected paths. Therefore, the interaction point in the NLoS hallway can be identified through the information about the coordinates of interaction positions from the simulation result.

On account of the time cost of channel measurement, the spatial resolution of the measurement is typically smaller than that of the simulation. Therefore, thanks to the accuracy of simulated dominant MPCs, the position of interaction points can be retraced more precisely by the simulation.
In the NLoS hallway, the LoS ray is absent, and only single- and multi-bounce reflected paths are left. However, in the simulation result, only near-NLoS Rx captures multi-bounce reflected paths, and the accuracy of the simulation decreases when the Rx position goes into the far-NLoS region. Both phenomena indicate that the derivation of major MPCs through RT is problematic at far-NLoS positions, and motives the research on how major MPCs evolve as the NLoS Rx moves. \rrev{In specific, we proposed a hybrid approach to derive major MPCs at far-NLoS positions by using the deterministic RT results at near-NLoS positions and deducing those at far-NLoS positions with a statistical evolving model. The derivation of non-major MPCs remains statistical.}

\subsection{Statistical Part based on Measurement}
\rev{In light of the accuracy of major MPCs from RT at near-NLoS positions and the problem in derivation of major MPCs through RT at far-NLoS positions, we propose a statistical evolving model to deduce major MPCs at far-NLoS positions from those at near-NLoS positions.} \rrev{In specific, the evolution of arrival angle, power, and delay of major MPCs are first characterized. Then statistical characteristics of MPCs that are not major are also summarized in this part.}

According to the observation on the continuous change of clusters in the spatial domain in Section~\ref{section: results_mpc}, we first investigate the evolution of interaction points in the NLoS hallway, \rrev{which is further converted into the arrival angle of major MPCs at far-NLoS positions.} The model is based on the position of the near-NLoS Rx as the reference and the corresponding initial interaction point obtained from RT and fitted by the measurement results given in Section~\ref{section: results_mpc}.

\begin{figure*}
    \centering
    \subfigure[\rrev{The geometrical relationship.}]
    {\includegraphics[width=0.3\linewidth]{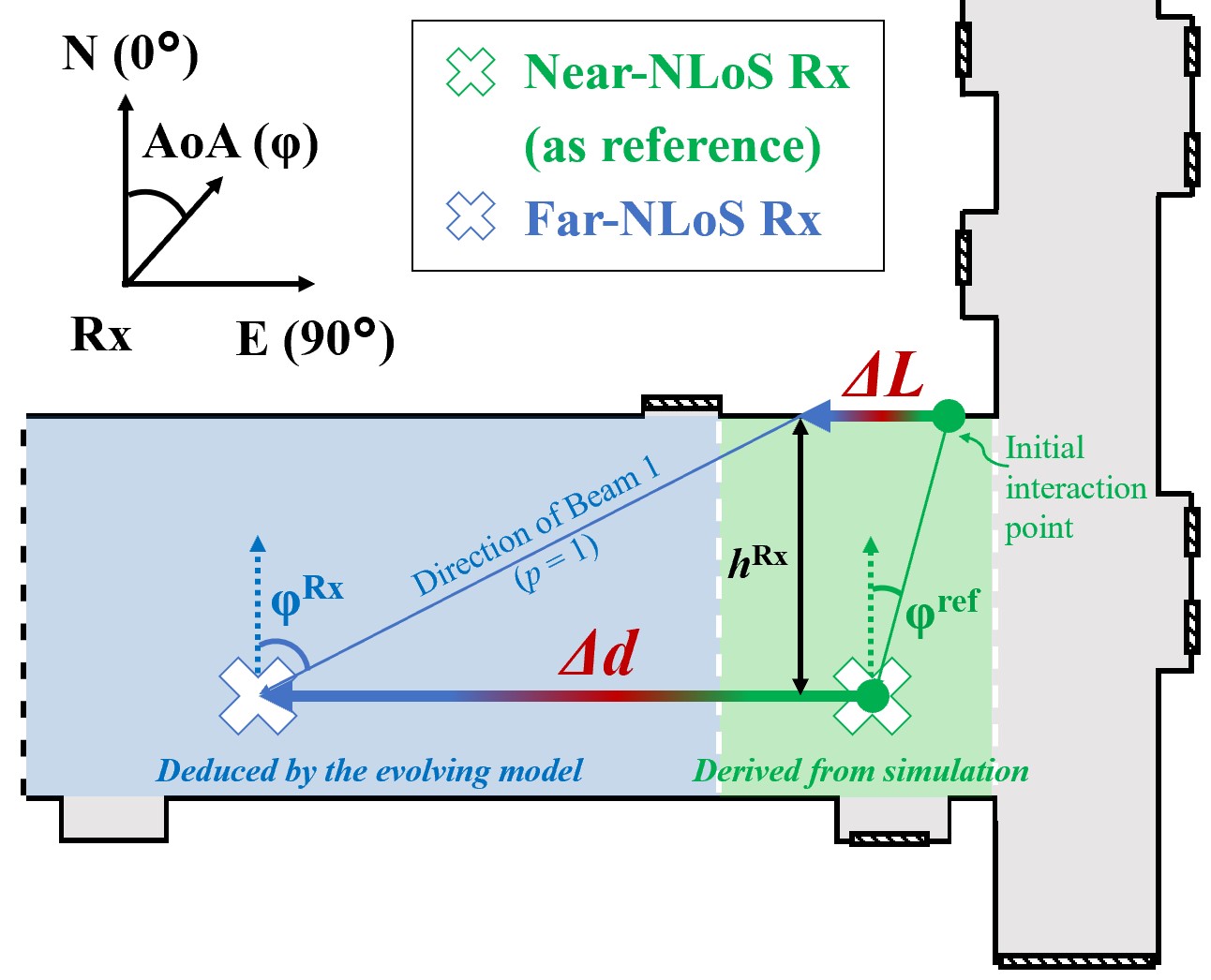}}
    \subfigure[Fitting result of the first row.]{\includegraphics[width=0.33\linewidth]{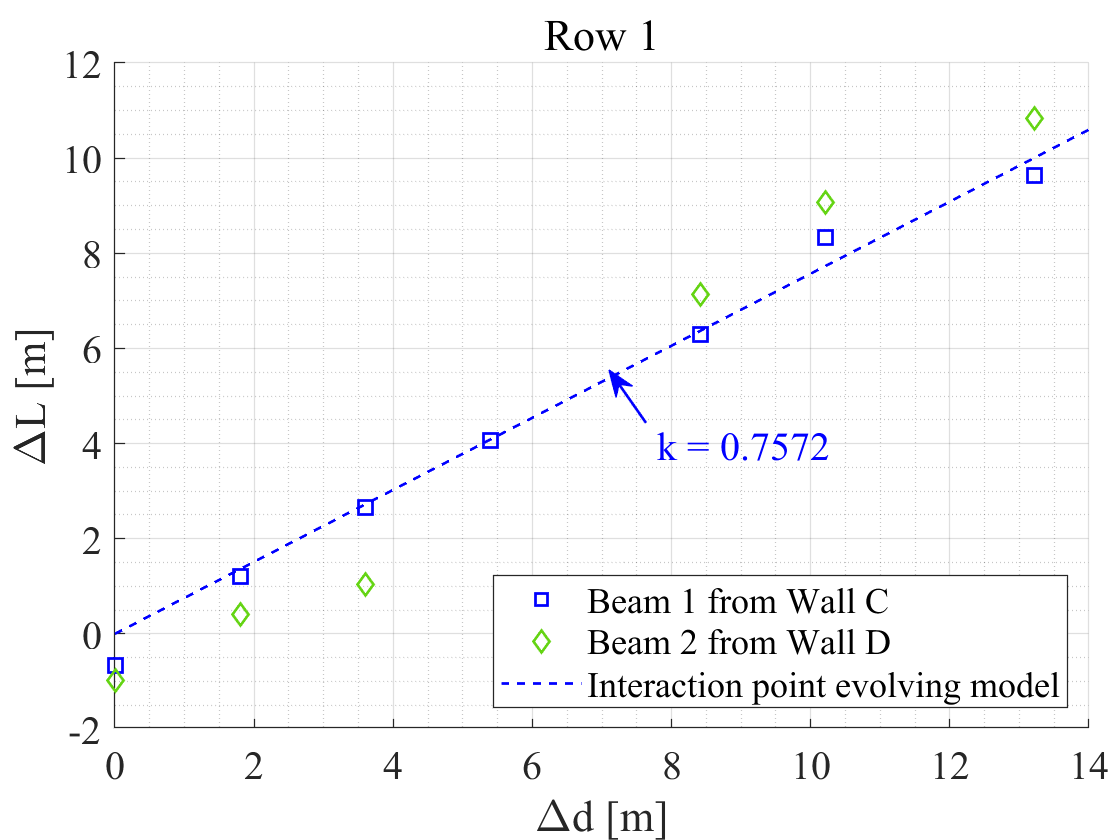}}
    \subfigure[Fitting result of the second row.]{\includegraphics[width=0.33\linewidth]{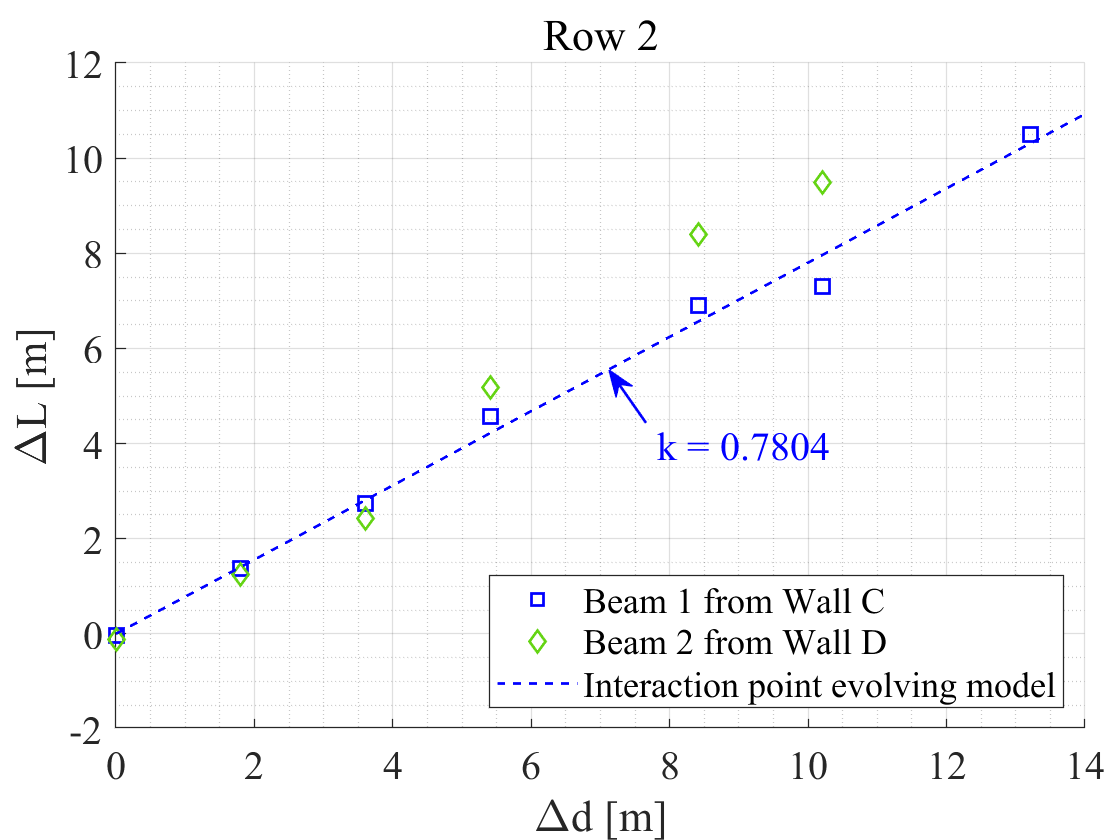}}
    \caption{The interaction point evolving model.}
    \label{fig:IO_evolving_fitting}
\end{figure*}
As illustrated in Fig.~\ref{fig:IO_evolving_fitting}(a), $\Delta d = d_{\rm Rx}-d_{\rm ref}$ denotes the distance between the far-NLoS Rx and the reference near-NLoS Rx in the same row, while $\Delta L$ represents the distance between the interaction point at the far-NLoS Rx and the initial interaction point. The knowledge of the interaction point can be converted to that of the arrival angle ($\varphi$, $\theta$) of the major MPC through geometric analysis. In the measurement result, the beam center is regarded as the arrival angle of the major MPC.
The analytic expression of $\Delta L$ with respect to $\Delta d$ is named as the interaction point evolving model. Fig.~\ref{fig:IO_evolving_fitting}(b)(c) shows the fitting result for two primary beams in each row in the NLoS hallway. Beam~1 with the AoA in the range of [$0^\circ$, $90^\circ$) is received from Wall~C, whose result shows a good linear fit between $\Delta L$ and $\Delta d$. By the definition of $\Delta L$ and $\Delta d$, the intercept of the interaction point evolving model, given by the linear fit, is fixed to zero. Beam~2 with the AoA in the range of ($90^\circ$, $180^\circ$] is received from Wall~D, whose result is affected by the indented office with a width of 0.8~m. By contrast, the office door on Wall~C is indented by 0.18~m, and thus Beam~1 is less influenced. The fitting results indicate that the relation between $\Delta d$ and $\Delta L$ can be modeled by a linear fit through the origin, as
\begin{equation}
    \Delta L = k \cdot \Delta d,
\end{equation}
where the empirical value of the slope $k$ is about 0.7-0.8 in the NLoS hallway.

\begin{figure}
    \centering
    \subfigure[The first row.]{\includegraphics[width=\linewidth]{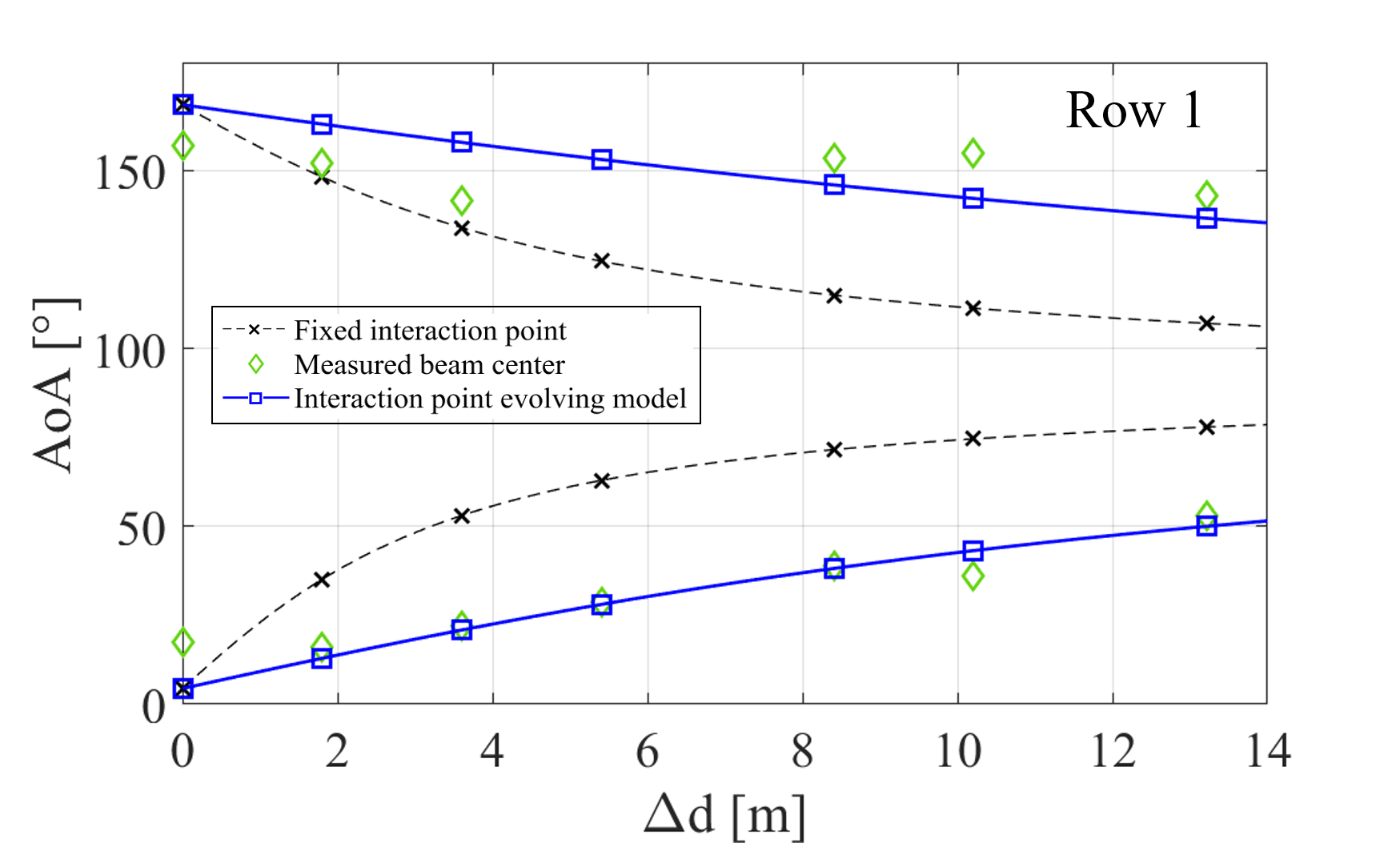}}
    \\
    \subfigure[The second row.]{\includegraphics[width=\linewidth]{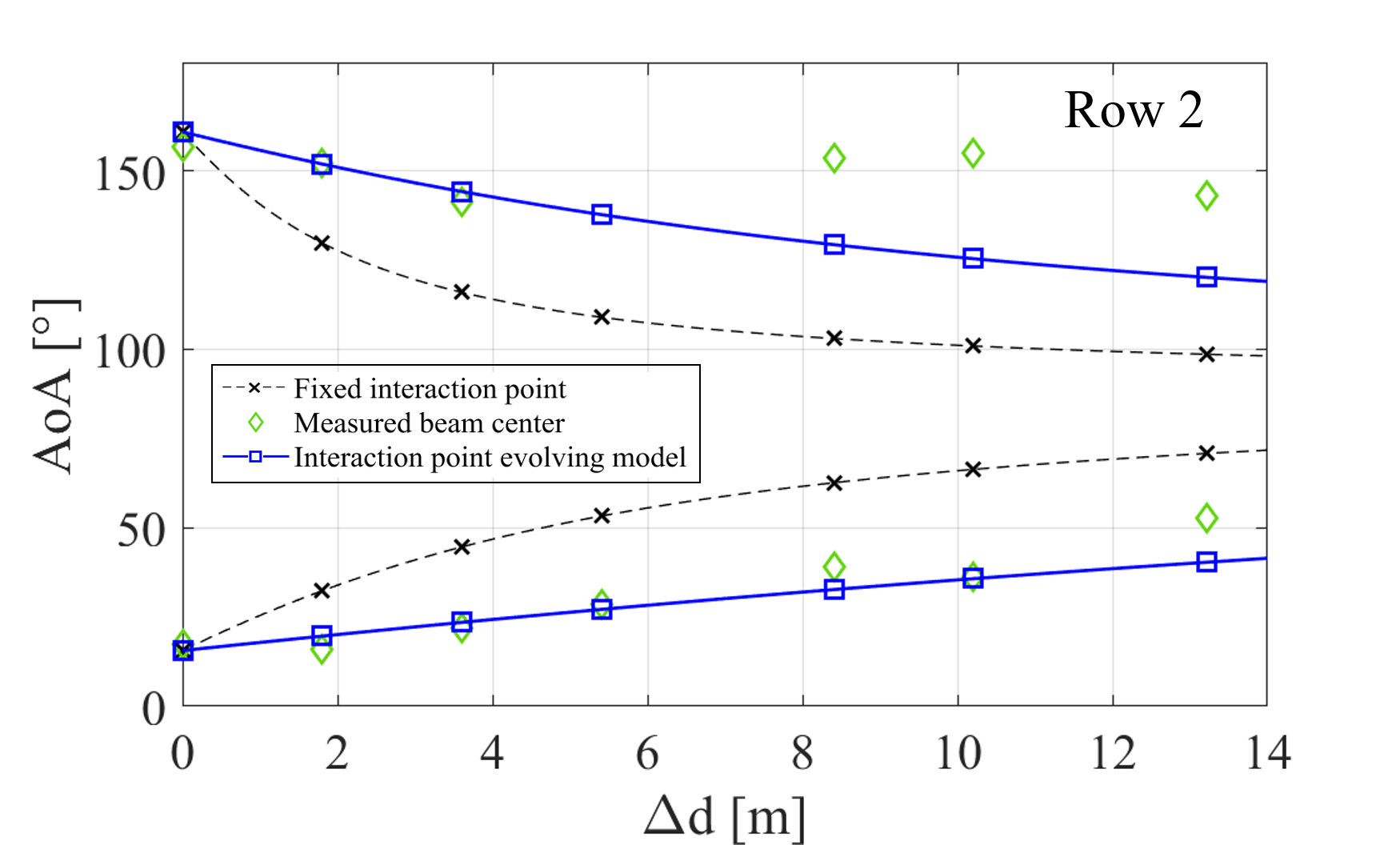}}
    \caption{The AoA of beam corresponding to the distance between the NLoS Rx and the reference near-NLoS Rx ($\Delta d$).}
    \label{fig:IO_evolving}
\end{figure}

\begin{figure*}
    \centering
    \includegraphics[width=\linewidth]{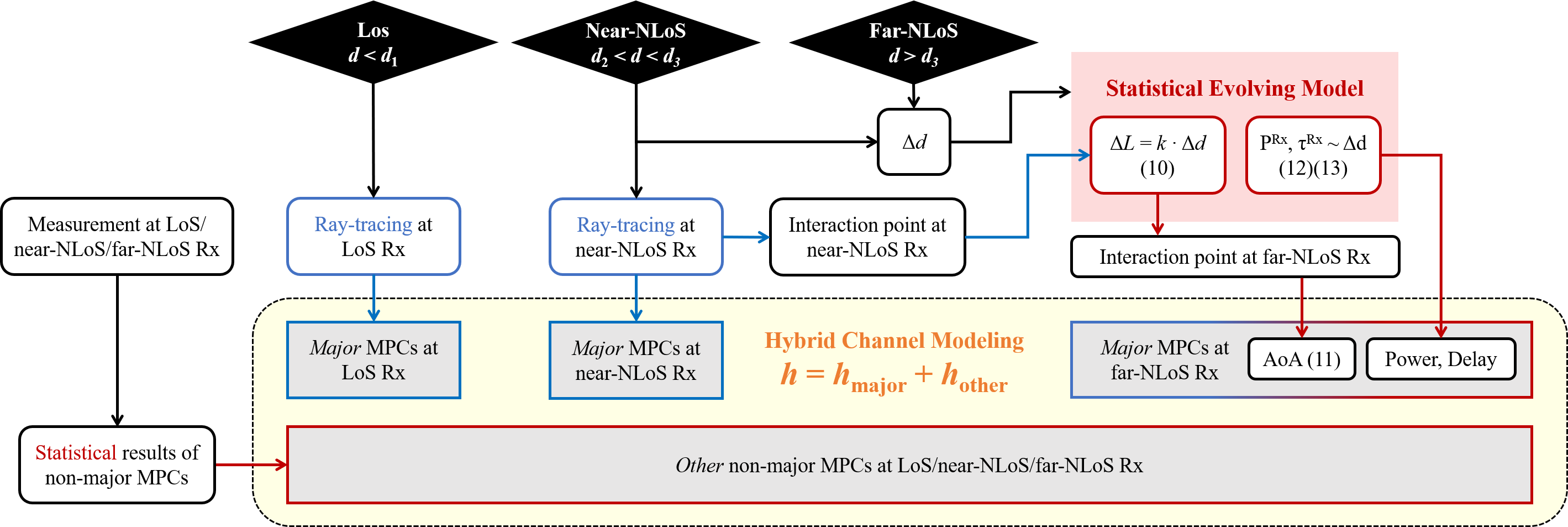}
    \caption{\rrev{Flow chart of the hybrid channel modeling.}}
    \label{fig:model_method}
\end{figure*}

\rrev{As illustrated in Fig.~\ref{fig:model_method}, by applying the simulation result at the reference near-NLoS Rx into the interaction point evolving model, the coordinates of interaction points at far-NLoS Rx in the NLoS hallway can be determined. With the knowledge of the interaction point and the location of the far-NLoS Rx, the arrival angle of major MPCs is derived,} i.e., $\angle$(IO,Rx)$_{p}$ = ($\varphi_p^{\rm Rx}$, $\theta_p^{\rm Rx}$), $p=1,2,...,n^{\rm Rx}$ where $n^{\rm Rx}$ is the number of major MPCs at each Rx. \rev{In this scenario, we find two major MPCs ($n^{\rm Rx}=2$), one from Wall~C ($p=1$) and the other from Wall~D ($p=2$).} As the elevation domain is not fully explored in our measurement due to the time cost, we focus on the horizontal property, i.e., the azimuth angle-of-arrival (AoA, $\varphi$) of major MPCs at each NLoS Rx, which can be calculated by
\begin{equation}\label{eq:drift}
    \varphi_{p}^{\rm Rx} [^\circ] =
    \begin{cases}
    90 - \arctan\left(\frac{h^{\rm Rx}}{\Delta d-\Delta L+h^{\rm Rx} \cdot\tan(\varphi_{p}^{\rm ref})}\right) & p=1 \\
    90 + \arctan\left(\frac{h^{\rm Rx}}{\Delta d-\Delta L+h^{\rm Rx} \cdot\tan(\varphi_{p}^{\rm ref})}\right) & p=2
    \end{cases}
\end{equation}
where $h^{\rm Rx}$ is the distance of the Rx to Wall~C. \rev{$\varphi_{p}^{\rm ref}$ is the AoA, corresponding to the initial interaction point, of the $p$-th major MPC at the reference Rx given by the simulation.
The result of major MPCs' AoA at NLoS Rx is shown in Fig.~\ref{fig:IO_evolving}. The black dotted line is a benchmark representing the AoA of major MPCs at each NLoS Rx by assuming $\Delta L=0$, i.e. all NLoS Rx share the same, fixed interaction point as the reference Rx. The result under this assumption deviates from the AoAs from the measurement (green diamonds), and confirms that the interaction point of the major MPC is drifting away from the initial interaction point. The drifting of interaction points is better characterized by the evolving model (blue line) as \eqref{eq:drift}, especially for major MPCs from Wall~C with less indented offices.}

\begin{figure}
    \centering
    \subfigure[Power of major MPCs.]{\includegraphics[width=\linewidth]{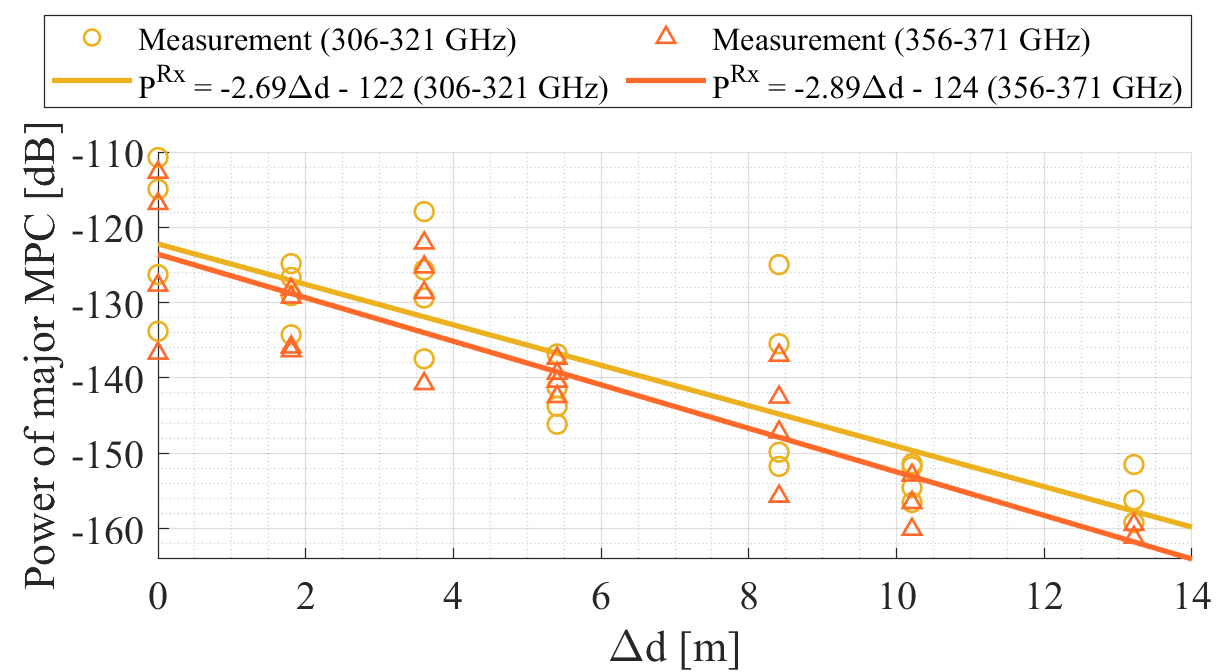}}
    \\
    \subfigure[Delay of major MPCs.]{\includegraphics[width=\linewidth]{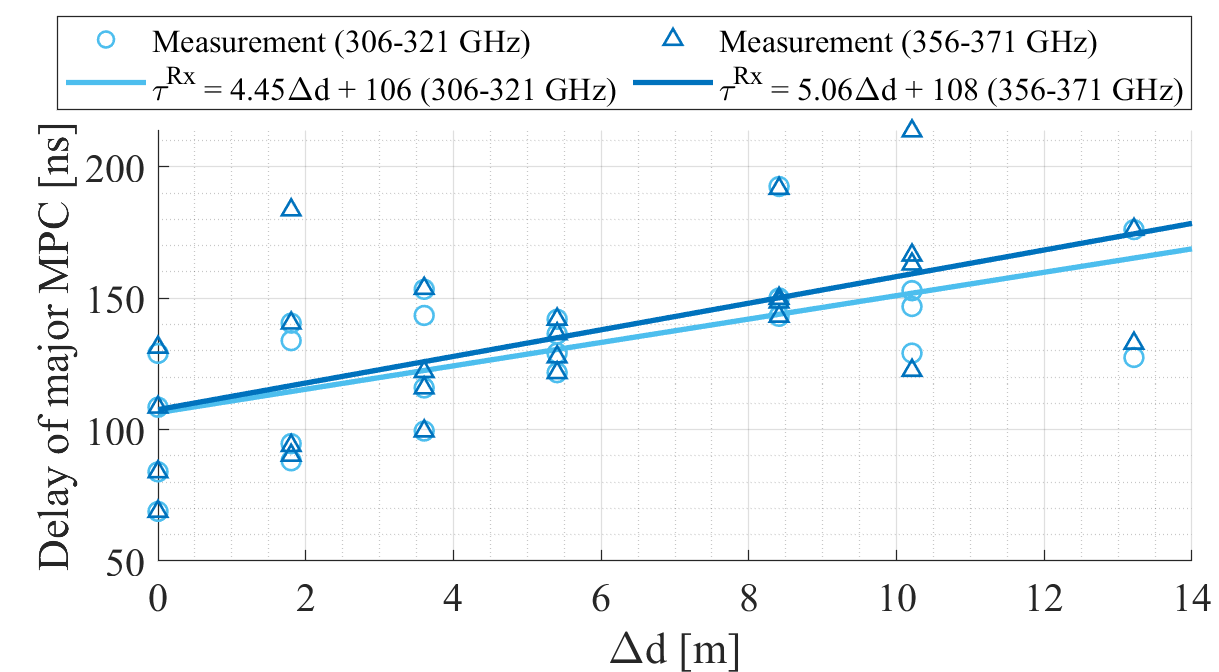}}
    \caption{\rev{Statistical results of major MPCs.}}
    \label{fig:power_delay_major_MPC}
\end{figure}

\rev{Following Fig.~\ref{fig:model_method}, we then characterize power and delay of major MPCs with respect to $\Delta d$. With the knowledge of arrival angles ($\varphi_p^{\rm Rx}$, $\theta_p^{\rm Rx}$), major MPCs in the measurement data are determined by searching for the MPC with the maximum power in clusters which contain MPCs from the angles. Then, the power and delay of major MPCs can be statistically characterized.} Fig.~\ref{fig:power_delay_major_MPC} shows that the power (in dB) and the delay (in ns) of major MPCs are respectively linearly related to $\Delta d$, the distance with respect to the reference Rx. \rev{In specific, the fitting results in dual bands are expressed in \eqref{eq:power_fit_major_mpc} and \eqref{eq:delay_fit_major_mpc}. At 356-371~GHz, the power of major MPCs is smaller and the delay is larger, compared with counterparts at 306-321~GHz. Moreover, the offsets are close in dual bands while the slopes are steeper at the higher frequency band.
\begin{equation}\label{eq:power_fit_major_mpc}
    P^{\rm Rx} \text{ [dB]} =
    \begin{cases}
        -2.69\cdot\Delta d - 122, & \text{306-321~GHz}, \\
        -2.89\cdot\Delta d - 124, & \text{356-371~GHz}. \\
    \end{cases}
\end{equation}
\begin{equation}\label{eq:delay_fit_major_mpc}
    \tau^{\rm Rx} \text{ [ns]} =
    \begin{cases}
        4.45\cdot\Delta d + 106, & \text{306-321~GHz}, \\
        5.06\cdot\Delta d + 108, & \text{356-371~GHz}. \\
    \end{cases}
\end{equation}}

So far we have characterized the evolution of major MPCs, including power, delay, and arrival angle, at NLoS positions. The idea is summarized in Fig.~\ref{fig:model_method}. \rrev{In short, for far-NLoS Rx, determination of major MPCs does not rely on RT. Instead, it reuses the deterministic RT result at near-NLoS positions and the statistical evolving model.} This method reduces the computational cost of the deterministic part, and simultaneously solves the problem in the derivation of major MPCs through RT at far-NLoS positions.

\begin{figure}
    \centering
    \subfigure[Number of subpaths.]
    {\includegraphics[width=\linewidth]{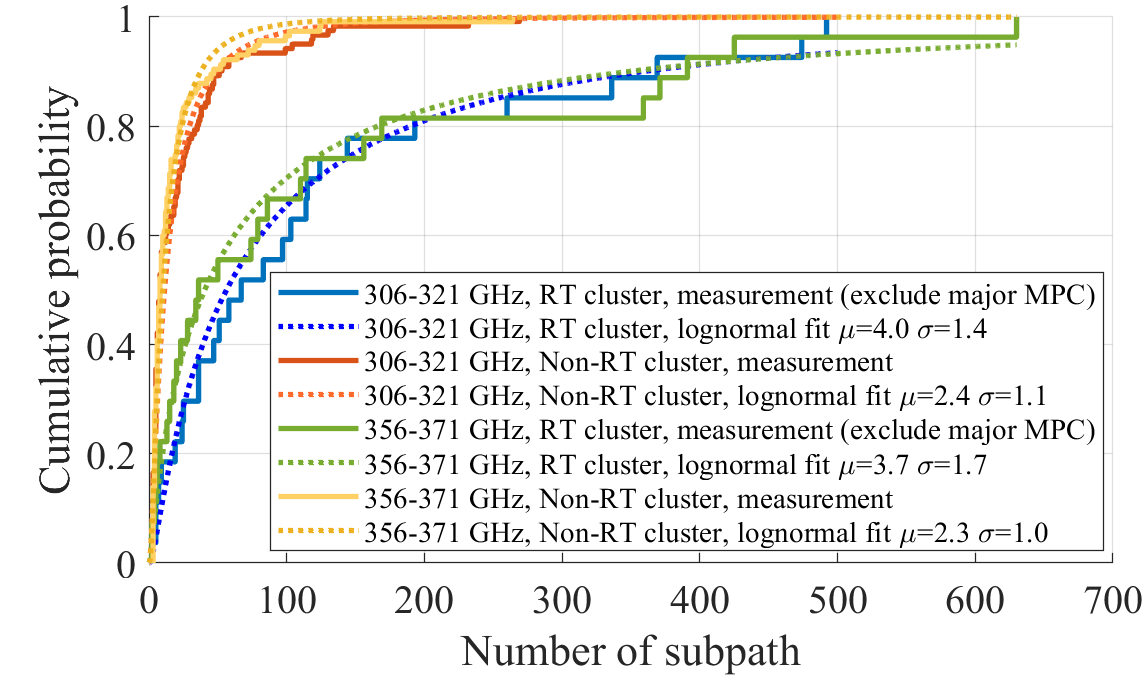}}
    \\
    \subfigure[Intra-cluster delay spread.]{\includegraphics[width=\linewidth]{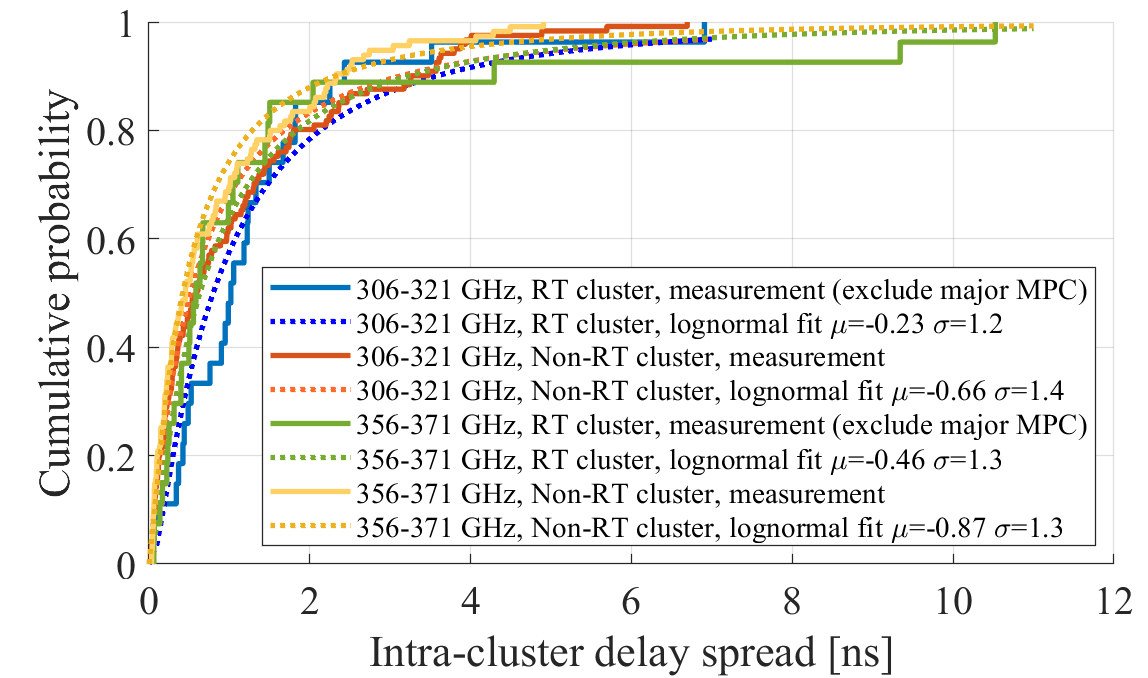}}
    \\
    \subfigure[Intra-cluster azimuth spread of angle.]{\includegraphics[width=\linewidth]{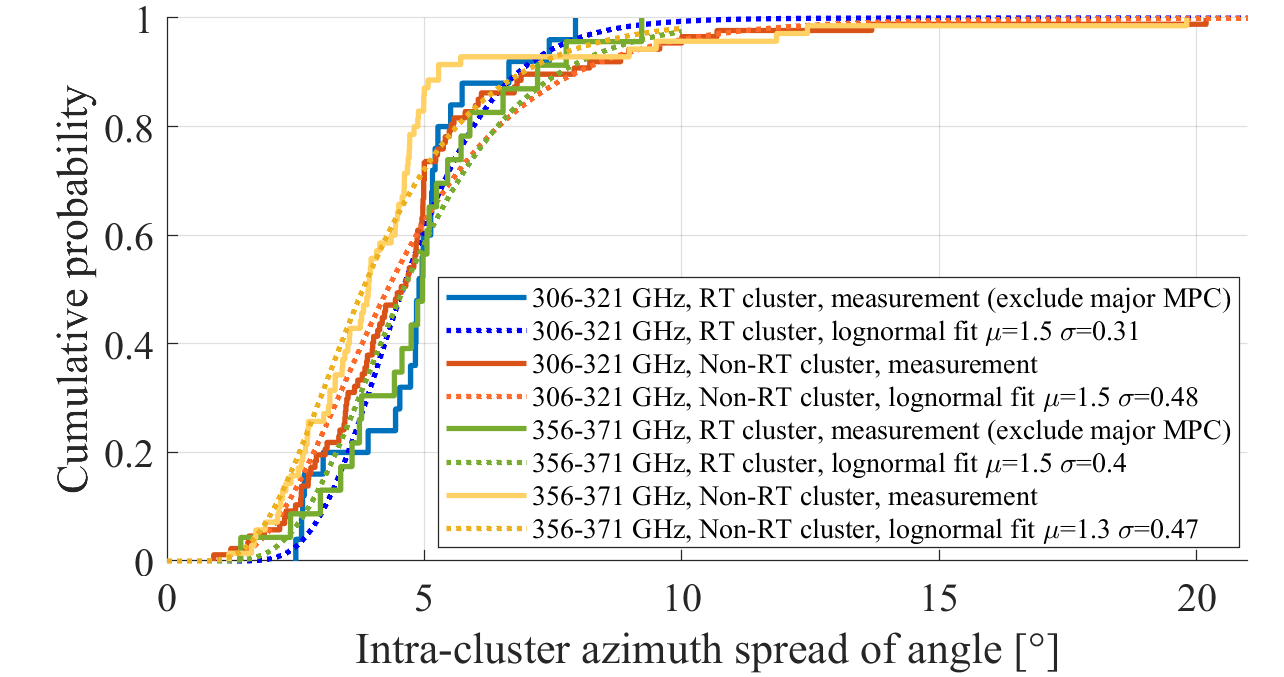}}
    \caption{\rev{Statistical results of RT and non-RT clusters.}}
    \label{fig:statistic_result}
\end{figure}

After determining major MPCs, the clusters can be classified into RT clusters, which contains major MPCs, and otherwise non-RT clusters. Since subpaths in the RT clusters except the major MPC and those in non-RT clusters can be statistically generated. We hereby statistically characterize the number of subpaths, and intra-cluster delay and angular spreads of non-RT clusters and RT clusters excluding the major MPC. Fig.~\ref{fig:statistic_result} shows that these parameters follow a log-normal distribution, respectively. \rev{The characteristics are found statistical-wise the same and parameter-wise different between dual bands.}
\section{Conclusion} \label{section: conclusion}
\rev{In this paper, we have conducted channel measurement, simulation, and RT-statistical hybrid channel modeling for the NLoS region in an L-shaped scenario.
First, we carried out a dual-band angular-resolvable wideband channel measurement in an indoor L-shaped hallway at 306-321~GHz and 356-371~GHz. The complete analysis and comparison of THz channel characteristics in dual bands is summarized in Table~\ref{tab:characteristic_noncluster}. The characteristics are found statistical-wise the same and parameter-wise different between dual bands. The frequency band of 356-371~GHz has higher path loss and smaller ASA and DS in general compared with the counterparts at 306-321~GHz.
Moreover, in light of the measurement result, we proposed a modified $\alpha-\beta$ path loss model, which is verified in the NLoS region for both indoor and outdoor L-shaped scenarios.
Besides, an RT-based simulation in the L-shaped hallway is implemented. The simulated major MPCs, i.e., multi-bounce reflected paths, is verified accurate at near-NLoS positions, whereas the derivation of major MPCs through RT at far-NLoS positions is problematic. The phenomenon further motivates the modeling of major MPC evolution in the NLoS region of the L-shaped scenario.
The proposed evolving model takes the RT result at near-NLoS positions as reference and deduces AoA, power, and delay of major MPCs at far-NLoS positions. In particular, the interaction point on the wall is drifting from the initial position by $\Delta L$, which is propositional to the distance between this Rx and the near-NLoS position, $\Delta d$. The azimuth angle of arrival can then be calculated through geometric analysis. The power and delay of major MPCs at NLoS Rx can be modeled by a linear fit with respect to $\Delta d$.
Finally, an RT-statistical hybrid channel modeling method is tailored for the THz NLoS hallway scenario. The deterministic process in hybrid channel modeling uses RT modeling of dominant MPCs, i.e., multi-bounce reflected paths, in the near-NLoS region, while dominant MPCs at other NLoS positions can be deduced based on the statistical evolving model. The proposed approach remedies the problem in the derivation of dominant MPCs through RT at far-NLoS positions and reduces the computational cost of channel modeling in the NLoS region.}



\bibliographystyle{IEEEtran}
\bibliography{bibliography}

\end{document}